\documentclass[sigconf,edbt]{acmart-edbt2024}

\def\BibTeX{{\rm B\kern-.05em{\sc i\kern-.025em b}\kern-.08em
    T\kern-.1667em\lower.7ex\hbox{E}\kern-.125emX}}

\usepackage{booktabs} 
\usepackage{balance}
\usepackage[ruled, vlined, linesnumbered, longend]{algorithm2e}
\usepackage{subcaption}
\usepackage{utfsym}
\usepackage{paralist}
\usepackage{tikz}
\usetikzlibrary{trees}
\usetikzlibrary{calc,arrows.meta,positioning}
\usetikzlibrary{patterns,patterns.meta} 
\usepackage[a-1b]{pdfx}

\newcommand{\spara}[1]{\smallskip\noindent{\textbf{#1}}}

\newcommand{\sdot}{\ensuremath{.\,}}

\newcommand{\zindex}{{Z-index}\xspace}

\newcommand{\zcurve}{{Z-curve}\xspace}

\newcommand{\bigo}[1]{\ensuremath{O(#1)}}

\newcommand{\twodim}{2D\xspace}

\renewcommand{\aa}{\ensuremath{a}\xspace}
\newcommand{\bb}{\ensuremath{b}\xspace}

\newcommand{\qdistr}{\ensuremath{\mathcal{Q}}\xspace}
\newcommand{\ddistr}{\ensuremath{\mathcal{D}}\xspace}
\newcommand{\abcd}{\ensuremath{\textsc{abcd}}\xspace}
\newcommand{\acbd}{\ensuremath{\textsc{acbd}}\xspace}

\newcommand{\normcost}[1]{\ensuremath{{cost}_{{#1}}}\xspace}
\newcommand{\totalcost}[1]{\ensuremath{\mathcal{C}_{#1}}\xspace}
\newcommand{\costredemp}[1]{\ensuremath{{red}_{{#1}}}\xspace}

\newcommand{\zindexinstance}{\ensuremath{\mathcal{Z}}\xspace}

\newcommand{\cellX}{\ensuremath{X}\xspace}
\newcommand{\cellY}{\ensuremath{Y}\xspace}
\newcommand{\A}{{\textsc{A}}\xspace}
\newcommand{\B}{{\textsc{B}}\xspace}
\newcommand{\CC}{{\textsc{C}}\xspace}
\newcommand{\D}{{\textsc{D}}\xspace}

\newcommand{\xx}{\ensuremath{x}}
\newcommand{\yy}{\ensuremath{y}}
\newcommand{\xval}{\xx\xspace}
\newcommand{\yval}{\yy\xspace}
\newcommand{\xopt}{\ensuremath{\mathbf{x}}\xspace}
\newcommand{\yopt}{\ensuremath{\mathbf{y}}\xspace}
\newcommand{\partitionpoint}{(\xopt,\yopt)\xspace}

\newcommand{\qq}[2]{\ensuremath{{q}_{_{{#1}{#2}}}}\xspace}
\newcommand{\nn}[1]{\ensuremath{{n}_{_{{#1}}}}\xspace}
\newcommand{\numdata}{\ensuremath{{N}}\xspace}

\newcommand{\cellrange}[3]{\ensuremath{{#1}\in{#2}{#3}}}
\newcommand{\iden}[2]{\ensuremath{\textsc{$\delta$}_{\cellrange{\query}{#1}{#2}}}\xspace}

\newcommand{\pagefactor}{\ensuremath{\alpha}\xspace}

\newcommand{\leafsize}{\ensuremath{L}\xspace}
\newcommand{\pagelist}{\textsc{PageList}\xspace}
\newcommand{\leaflist}{\textsc{LeafList}\xspace}
\newcommand{\page}{\ensuremath{P}\xspace}

\newcommand{\mbr}{\ensuremath{bbs}\xspace}

\newcommand{\query}{\ensuremath{{\textsc{R}}}\xspace}
\newcommand{\bottomleft}{\ensuremath{\textsc{BL}}\xspace}
\newcommand{\topright}{\ensuremath{\textsc{TR}}\xspace}

\newcommand{\querypoint}{\ensuremath{\mathbf{p}}\xspace}

\newcommand{\filter}{\textsc{Filter}\xspace}

\newcommand{\bool}{\textsc{b}\xspace}
\newcommand{\treetraversal}{\textsc{TreeTraversal}\xspace}
\newcommand{\rangequery}{\textsc{Range-query}\xspace}

\newcommand{\pagequeryoverlap}{\textsc{PageQueryOverlap}\xspace}
\newcommand{\overlapresult}{\ensuremath{overlap}\xspace}
\newcommand{\nextpage}{\textsc{NextPage}\xspace}

\newcommand{\low}{\textsc{low}\xspace}
\newcommand{\high}{\textsc{high}\xspace}

\newcommand{\qbelow}{\textsc{Below}\xspace}
\newcommand{\qabove}{\textsc{Above}\xspace}
\newcommand{\qleft}{\textsc{Left}\xspace}
\newcommand{\qright}{\textsc{Right}\xspace}
\newcommand{\qcase}{\textsc{case}\xspace}

\newcommand{\queryresult}{\ensuremath{result}\xspace}

\newcommand{\rank}{\ensuremath{\textsc{Ord}}\xspace}
\newcommand{\orderValue}{\ensuremath{\textnormal{o}}\xspace}
\newcommand{\optorderValue}{\ensuremath{\mathbf{o}}\xspace}
\newcommand{\hierPart}{\ensuremath{\mathbf{h}}\xspace}

\newcommand{\greedy}{\textsc{Greedy}\xspace}

\newcommand{\waz}{\textsc{WaZI}\xspace}
\newcommand{\wazminus}{\textsc{WaZI-SK}\xspace}
\newcommand{\baseplus}{\textsc{Base+SK}\xspace}

\newcommand{\base}{\textsc{Base}\xspace}
\newcommand{\zpgm}{\textsc{Zpgm}\xspace}
\newcommand{\rsmi}{\textsc{Rsmi}\xspace}

\newcommand{\hrr}{\textsc{HRR}\xspace}
\newcommand{\str}{\textsc{STR}\xspace}
\newcommand{\cur}{\textsc{CUR}\xspace}
\newcommand{\quasii}{\textsc{QUASII}\xspace}
\newcommand{\bmtree}{\textsc{BMTree}\xspace}
\newcommand{\lmsfc}{\textsc{LMSFC}\xspace}

\newcommand{\flood}{\textsc{Flood}\xspace}
\newcommand{\quilts}{\textsc{QUILTS}\xspace}

\newcommand{\qdtreegreedy}{\textsc{Qd-Gr}\xspace}

\newcommand{\numsamples}{\ensuremath{\kappa}\xspace}

\newcommand{\calinev}{{\sc{CaliNev}}\xspace}
\newcommand{\japan}{{\sc{Japan}}\xspace}
\newcommand{\nycity}{{\sc{NewYork}}\xspace}
\newcommand{\iberia}{{\sc{Iberia}}\xspace}

\newcommand{\coderepo}{\url{https://version.helsinki.fi/ads/learned-zindex}\xspace}

\newcommand{\greentick}{\textcolor{green}{\usym{2714}}}
\newcommand{\redcross}{\textcolor{red}{\usym{2717}}}

\tikzset{onslide/.code args={<#1>#2}{%
  \only<#1>{\pgfkeysalso{#2}} 
}}
\tikzset{alt/.code args={<#1>#2#3}{%
  \alt<#1>{\pgfkeysalso{#2}}{\pgfkeysalso{#3}} 
}}
\tikzset{temporal/.code args={<#1>#2#3#4}{%
  \temporal<#1>{\pgfkeysalso{#2}}{\pgfkeysalso{#3}}{\pgfkeysalso{#4}} 
}}

\tikzset{
        position/.style args={#1:#2 from #3}{
            at=(#3.#1), anchor=#1+180, shift=(#1:#2)
        }
    }

\tikzset{size1/.style={scale=5}}
\tikzset{size2/.style={scale=4}}
\tikzset{size3/.style={scale=2}}

\tikzset{arrowstyle_thick/.style={draw,->,rounded corners,line width=3pt}}
\tikzset{arrowstyle_thin/.style={draw,->,rounded corners,line width=1.5pt}}
\tikzset{arrowstyle_verythin/.style={draw,->,rounded corners,line width=1pt}}

\tikzset{datapoint/.style={circle,inner sep=0mm,minimum size=0.2mm,draw,blue,fill=white,size1}}

\usepackage{xcolor}
\usepackage{colortbl}

\setcopyright{rightsretained}
\copyrightyear{2024}

\acmDOI{}

\acmISBN{978-3-89318-095-0}

\acmConference[EDBT 2024]{27th International Conference on Extending Database Technology (EDBT)}{25th March-28th March, 2024}{Paestum, Italy}
\acmYear{2024}

\begin{document}
\title{\textsc{WaZI}: A Learned and Workload-aware Z-Index}

\author{Sachith Pai}
\affiliation{%
  \institution{University of Helsinki}
  \city{Helsinki}
  \country{Finland}
}
\email{sachith.pai@helsinki.fi}

\author{Michael Mathioudakis}
\orcid{0000-0003-0074-3966}
\affiliation{%
  \institution{University of Helsinki}
  \city{Helsinki}
  \country{Finland}
}
\email{michael.mathioudakis@helsinki.fi}

\author{Yanhao Wang}
\orcid{0000-0002-7661-3917}
\affiliation{%
  \institution{East China Normal University}
  \city{Shanghai}
  \country{China}
}
\email{yhwang@dase.ecnu.edu.cn}

\begin{abstract}
  Learned indexes fit machine learning (ML) models to the data and use them to make query operations more time and space-efficient. 
  Recent works propose using learned spatial indexes to improve spatial query performance by optimizing the storage layout or internal search structures according to the data distribution. 
  However, only a few learned indexes exploit the query workload distribution to enhance their performance. 
  In addition, building and updating learned spatial indexes are often costly on large datasets due to the inefficiency of (re)training ML models.
  
  In this paper, we present \waz, a learned and workload-aware variant of the \zindex, which jointly optimizes the storage layout and search structures, as a viable solution for the above challenges of spatial indexing.
  Specifically, we first formulate a cost function to measure the performance of a \zindex on a dataset for a range-query workload.
  Then, we optimize the \zindex structure by minimizing the cost function through adaptive partitioning and ordering for index construction.
  Moreover, we design a novel page-skipping mechanism to improve the query performance of \waz by reducing access to irrelevant data pages.
  Our extensive experiments show that the \waz index improves range query time by 40\% on average over the baselines while always performing better or comparably to state-of-the-art spatial indexes.
  Additionally, it also maintains good point query performance.
  Generally, \waz provides favorable tradeoffs among query latency, construction time, and index size.
\end{abstract}

\maketitle

\section{Introduction}
\label{sec:intro}

\begin{figure*}[t]
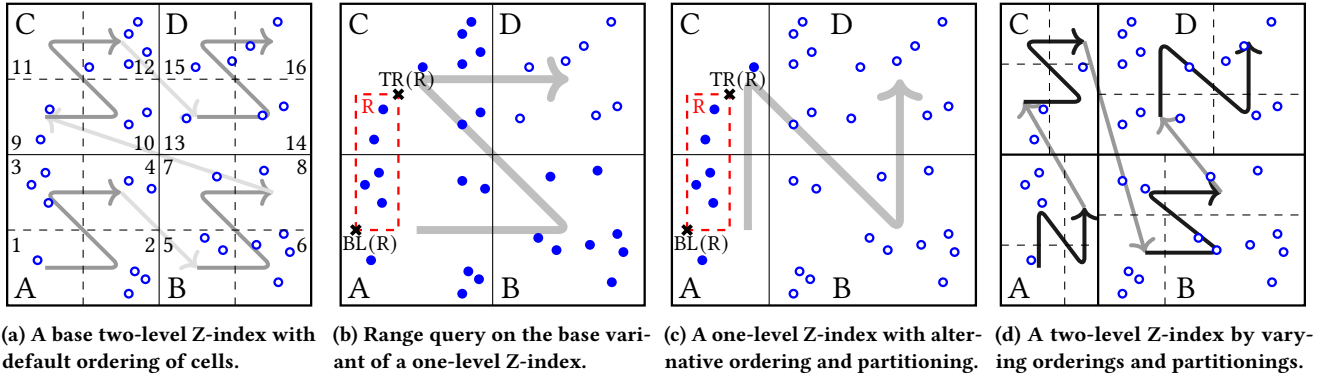

    \centering
    \begin{subfigure}{0.24\textwidth}
    \centering
    \tikz[line width=0.3mm,scale=4.0]{\path[arrowstyle_thin,black!40] (0.125,0.125) -- ++ (0.25,0) -- ++ (-0.25,0.25) -- ++ (0.25,0);
\path[arrowstyle_thin,gray!25] (0.375,0.375)  -- ++ (0.25,-0.25);
\path[arrowstyle_thin,black!40] (0.625,0.125) -- ++ (0.25,0) -- ++ (-0.25,0.25) -- ++ (0.25,0);
\path[arrowstyle_thin,gray!25] (0.875,0.375) -- ++ (-0.75,0.25);
\path[arrowstyle_thin,black!40] (0.125,0.625) -- ++ (0.25,0) -- ++ (-0.25,0.25) -- ++ (0.25,0);
\path[arrowstyle_thin,gray!25]  (0.375,0.875) -- ++ (0.25,-0.25);
\path[arrowstyle_thin,black!40] (0.625,0.625) -- ++ (0.25,0) -- ++ (-0.25,0.25) -- ++ (0.25,0);

\draw[black] (0,0) rectangle (1,1);
\tikzset{class1/.style={fill=white}}
\tikzset{class2/.style={fill=white}}

\input{./figures/tikz_figures/datapoints.tex}

\draw[thin] (0.5,0) -- (0.5,1);
\draw[thin] (0,0.5) -- (1,0.5);

\draw[thin,dashed] (0.25,0) -- (0.25,1);
\draw[thin,dashed] (0,0.25) -- (1,0.25);
\draw[thin,dashed] (0.75,0) -- (0.75,1);
\draw[thin,dashed] (0,0.75) -- (1,0.75);

\path (0.03,0.2) node[font=\normalsize]{1} ++ (0.44,0) node[font=\normalsize]{2} ++ (-0.44,0.26) node[font=\normalsize]{3} ++ (0.44,0)  node[font=\normalsize]{4} ++ (0.06,-0.26)node[font=\normalsize]{5} ++ (0.44,0) node[font=\normalsize]{6} ++ (-0.44,0.26) node[font=\normalsize]{7} ++ (0.44,0) node[font=\normalsize]{8} ++ (-0.94,0.08) node[font=\normalsize]{9} ++ (0.42,0) node[font=\normalsize]{10} ++ (-0.4,0.25) node[font=\normalsize]{11} ++ (0.4,0) node[font=\normalsize]{12} ++ (0.1,-0.25) node[font=\normalsize]{13} ++ (0.4,0) node[font=\normalsize]{14} ++ (-0.4,0.25) node[font=\normalsize]{15} ++ (0.4,0) node[font=\normalsize]{16};

\path (0.06,0.06) node[font=\LARGE] {A} ++ (0.5,0) node[font=\LARGE] {B}  ++ (0,0.875) node[font=\LARGE] {D}  ++ (-0.5,0) node[font=\LARGE] {C}; }
    \caption{A base two-level \zindex with default ordering of cells.}
    \label{fig:example:default}
    \end{subfigure}
    \hfill
    \begin{subfigure}{0.24\textwidth}
    \centering
    \tikz[line width=0.3mm,scale=4.0]{

\path[arrowstyle_thick,gray!50] (0.25,0.25) -- ++ (0.5,0) -- + (-0.5,0.5) -- + (0,0.5);

\draw[black] (0,0) rectangle (1,1);
\tikzset{class1/.style={fill=blue}}
\tikzset{class2/.style={fill=blue}}

\input{./figures/tikz_figures/datapoints.tex}

\draw[thin] (0.5,0) -- (0.5,1);
\draw[thin] (0,0.5) -- (1,0.5);


\path (0.06,0.06) node[font=\LARGE] {A} ++ (0.5,0) node[font=\LARGE] {B}  ++ (0,0.875) node[font=\LARGE] {D}  ++ (-0.5,0) node[font=\LARGE] {C};

\draw[red, dashed] (0.05,0.25) rectangle (0.19,0.7);

\draw[line width= 0.5mm ] (0.065,0.265) -- (0.035,0.235);
\draw[line width= 0.5mm ] (0.065,0.235) -- (0.035,0.265);

\draw[line width= 0.5mm ] (0.205,0.715) -- (0.175,0.685);
\draw[line width= 0.5mm ] (0.205,0.685) -- (0.175,0.715);

\path (0.1,0.203) node[font=\small] {$\bottomleft(\query)$} ++ (0.12,0.542) node[font=\small] {$\topright(\query)$};

\path (0.09,0.66) node[red, font=\small] {$\query$};}
    \caption{Range query on the base variant of a one-level \zindex.}
    \label{fig:example:defquery}
    \end{subfigure}
    \hfill
    \begin{subfigure}{0.24\textwidth}
    \centering
    \tikz[line width=0.3mm,scale=4.0]{

\path[arrowstyle_thick,gray!50] (0.25,0.25) -- ++ (0,0.5) -- + (0.5,-0.5) -- + (0.5,0);

\draw[black] (0,0) rectangle (1,1);
\tikzset{class1/.style={fill=blue}}
\tikzset{class2/.style={fill=white}}

\input{./figures/tikz_figures/datapoints.tex}

\draw[thin] (0.32,0) -- (0.32,1);
\draw[thin] (0,0.5) -- (1,0.5);

\path (0.06,0.06) node[font=\LARGE] {A} ++ (0.5,0) node[font=\LARGE] {B}  ++ (0,0.875) node[font=\LARGE] {D}  ++ (-0.5,0) node[font=\LARGE] {C};

\draw[red, dashed] (0.05,0.25) rectangle (0.19,0.7);

\draw[line width= 0.5mm ] (0.065,0.265) -- (0.035,0.235);
\draw[line width= 0.5mm ] (0.065,0.235) -- (0.035,0.265);

\draw[line width= 0.5mm ] (0.205,0.715) -- (0.175,0.685);
\draw[line width= 0.5mm ] (0.205,0.685) -- (0.175,0.715);

\path (0.1,0.203) node[font=\small] {$\bottomleft(\query)$} ++ (0.12,0.542) node[font=\small] {$\topright(\query)$};

\path (0.09,0.66) node[red, font=\small] {$\query$};}
    \caption{A one-level \zindex with alternative ordering and partitioning.}
    \label{fig:example:altsplitquery}
    \end{subfigure}
    \hfill
    \begin{subfigure}{0.24\textwidth}
    \centering
    \tikz[line width=0.3mm,scale=4.0]{

\path[arrowstyle_thin,black!90] (0.125,0.125) -- ++ (0,0.2) -- ++ (0.15,-0.2) -- ++ (0,0.2);
\path[arrowstyle_thin,gray!80] (0.275,0.325)  -- ++ (-0.2,0.35);
\path[arrowstyle_thin,black!90] (0.075,0.675) -- ++ (0.2,0) -- ++ (-0.2,0.2) -- ++ (0.2,0);
\path[arrowstyle_thin,gray!80] (0.275,0.875) -- ++ (0.2,-0.7);

\path[arrowstyle_thin,black!90] (0.475,0.175) -- ++ (0.25,0) -- ++ (-0.25,0.2) -- ++ (0.25,0);
\path[arrowstyle_thin,gray!80]  (0.725,0.375) -- ++ (-0.2,0.25);
\path[arrowstyle_thin,black!90] (0.525,0.625) -- ++ (0,0.25) -- ++ (0.29,-0.25) -- ++ (0,0.25);

\draw[black] (0,0) rectangle (1,1);

\tikzset{class1/.style={fill=white}}
\tikzset{class2/.style={fill=white}}

\input{./figures/tikz_figures/datapoints.tex}

\draw[] (0.32,0) -- (0.32,1);
\draw[] (0,0.5) -- (1,0.5);

\draw[thin,dashed] (0.21,0) -- (0.21,0.5);
\draw[thin,dashed] (0,0.2) -- (0.32,0.2);

\draw[thin,dashed] (0.54,0) -- (0.54,0.5);
\draw[thin,dashed] (0.32,0.3) -- (1,0.3);

\draw[thin,dashed] (0.16,0.5) -- (0.16,1);
\draw[thin,dashed] (0,0.8) -- (0.32,0.8);

\draw[thin,dashed] (0.7,0.5) -- (0.7,1);
\draw[thin,dashed] (0.32,0.7) -- (1,0.7);

\path (0.06,0.06) node[font=\LARGE] {A} ++ (0.55,0) node[font=\LARGE] {B}  ++ (0,0.875) node[font=\LARGE] {D}  ++ (-0.55,0) node[font=\LARGE] {C};

}
    \caption{A two-level \zindex by varying orderings and partitionings.}
    \label{fig:example:alt2levels}
    \end{subfigure}
    \caption{Illustrations of the base \zcurve and \zindex and their variants proposed in this work.}
    \label{fig:example}
\end{figure*}

Spatial query processing has attracted significant interest in the database community over the last three decades with the proliferation of location-based services (LBSs).
Although numerous index structures such as R-trees~\cite{DBLP:conf/sigmod/Guttman84} and k-d trees~\cite{DBLP:journals/cacm/Bentley75} have been deployed in database systems to improve spatial query performance, their efficiency still cannot fully satisfy the requirements of real-world applications due to the rapid growth in the volume of spatial data.
The seminal work of Kraska et al.~\cite{DBLP:conf/sigmod/KraskaBCDP18} inspired the introduction of machine learning (ML) based indexes (i.e., \emph{learned indexes}) \cite{DBLP:conf/sigmod/KraskaBCDP18,DBLP:journals/pvldb/FerraginaV20,DBLP:conf/sigmod/DingMYWDLZCGKLK20,DBLP:conf/edbt/0001H21,DBLP:conf/mdm/WangFX019,DBLP:journals/pvldb/QiLJK20,DBLP:conf/sigmod/Li0ZY020,DBLP:conf/ssd/ZhangRLZ21} to replace their traditional counterparts as a way of improving query performance by exploiting data or query patterns or both while reducing space usage.
Most of the learned indexes developed are specific for one-dimensional data~\cite{DBLP:conf/sigmod/KraskaBCDP18,DBLP:conf/sigmod/GalakatosMBFK19,DBLP:journals/pvldb/FerraginaV20,DBLP:conf/sigmod/DingMYWDLZCGKLK20,DBLP:conf/edbt/0001H21}, which follows the abstraction that an index is a model that predicts the position of an element in a sorted array.
Such an abstraction is made possible only by the prerequisite that the data resides in a sorted array.
Therefore, any model that accurately and efficiently learns the data's cumulative density function (CDF) can serve as an index.
However, the above abstraction is not squarely applicable to spatial indexes since there is no predefined ordering between data points. 
%
Nonetheless, a few learned spatial indexes have attempted to overcome this challenge by using space-filling curves (SFCs) to linearize the spatial indexing task \cite{DBLP:journals/tods/QiTCZ20,DBLP:conf/mdm/WangFX019,DBLP:journals/pvldb/QiLJK20}. 
The construction procedure for such indexes proceeds in two phases.
The first phase uses space-filling curves for data linearization.
The second phase performs the transformed one-dimensional indexing task.
These indexes usually optimize the second phase of the construction while using standard methods for the first phase (linearization), missing out on the opportunity to tailor the linearization to fit the data and workload better.
To remedy the problem, in this work, we present a generalized variant of the \zindex that is both data- and workload-aware.
A \zindex is an intuitively simple spatial index structure with a long history in database management systems~\cite{ramakrishnan2003database}.
It uses a \emph{\zcurve} (a.k.a. \emph{Z-order curve}) to compute a sort order, denoted by \rank, for multi-dimensional data points.
%
Figure~\ref{fig:example:default} shows an example of the \zcurve and the corresponding \zindex.
The \zcurve visits the data points according to a hierarchical partitioning of the data space into cells and a specific ordering of these cells.
For instance, at the top level, the space is partitioned into four cells, namely \A, \B, \CC, and \D; at the second level, each of these cells is partitioned into four sub-cells, and so on, with the partitioning happening at the coordinates corresponding to the median of the data along each axis.
Within each cell, the ordering of its sub-cells consistently follows the same `Z'-like pattern: cells are ordered as ``\abcd'' for the first level, with higher-level cells having higher priority in the point ordering.
Each cell is further partitioned and ordered until the cells reach a predetermined size \leafsize.

The \zindex has the desirable property of \emph{monotonicity}; that is, any point ``dominated'' (i.e., having smaller values in both dimensions) by point \aa in the data space will always have a smaller sort order \rank than \aa.
This monotonicity property facilitates the processing of range queries, represented as rectangles in the data space.
A standard range-query algorithm is to obtain the locations in \rank of the bottom-left and top-right points of the query rectangle, scan the data entries between the two, and filter the data points that satisfy the query range.
However, the efficiency of such a range-query algorithm could still be vastly improved.
For example, let us consider the range query with rectangle $R$ in Figure~\ref{fig:example:defquery}: the algorithm processes many data entries outside the result set (e.g., all points in cell \B), which could lead to high query latencies.

\spara{Our Contributions.}
We propose a generalization of the \zindex that adapts gracefully to both the distribution of spatial data and the workload of range queries,  mitigating the retrieval of redundant data entries, and thus leading to improved query latencies at little additional cost to index construction.
Specifically, we present a method to construct a learned and workload-aware space-filling curve to improve the range-query performance of a generalized \zindex variant.
The \underline{W}orkload-\underline{a}ware \underline{Z}-\underline{I}ndex (\waz) variant we propose is flexible in two ways compared to the base \zindex: the partition location and the ordering of child cells.
Intuitively, it pays off to partition the data space so that cells correspond to regions that are fetched by several similar range queries.
Such a partitioning allows the index to avoid processing points in other cells during the execution of these queries.
For example, the \zindex with an alternative ordering and partitioning shown in Figure~\ref{fig:example:altsplitquery} processes fewer redundant data entries than the base \zindex in Figure~\ref{fig:example:defquery}.
Therefore, if similar queries dominate the anticipated query workload, the alternative ordering and partitioning of cells can reduce query latency.
The \zindex variant we propose is general enough to allow different ordering and partitioning for each cell across the index hierarchy, as shown in Figure~\ref{fig:example:alt2levels}.
Additionally, by harnessing the structural properties of the index, we develop and incorporate an efficient mechanism to reduce the computation required by skipping non-overlapping pages during range query processing.

In the experiments, we compare \waz with four state-of-the-art spatial indexes including \flood \cite{DBLP:conf/sigmod/NathanDAK20}, \str \cite{DBLP:conf/icde/LeuteneggerEL97}, a query-aware incremental index \quasii \cite{DBLP:conf/edbt/PavlovicSHA18}, and a cost-optimized variant of R-tree \cur \cite{DBLP:conf/ssdbm/RossSS01}.
The baselines considered cover the scope of indexes regarding query-awareness and the use of learned components.
The results show that \waz performs better than these baselines in terms of range and point query latency while exhibiting favorable tradeoffs regarding construction time and index size.
In the critical metric of range query latency, \waz significantly outperforms all the baselines.
The main contributions of this paper include:
\begin{enumerate}[1)]
  \item formalizing the problem of optimizing the partitioning and Z-ordering for the given data and anticipated query workload;
  \item providing an index construction algorithm that minimizes the retrieval cost;
  \item presenting a mechanism to efficiently skip over large regions of irrelevant pages to avoid the computational cost associated with skipped pages;
  \item providing an experimental evaluation of the proposed index against existing approaches, addressing the cost redemption behavior and effects of workload changes, as well as an ablation study to demonstrate the benefit from different components of our approach.
\end{enumerate}
An extended abstract associated with this work was presented at a non-archival workshop~\cite{aidb22}.

\section{Related Work}
\label{sec:literature}

In this section, we review the existing literature on \emph{learned indexes} and \emph{spatial indexes}, which are closely relevant to the problem we study in this paper.

\spara{Traditional Spatial Indexes.}
Spatial indexes~\cite{DBLP:journals/csur/GaedeG98} were well studied across several decades in the database community.
Existing (non-learned) spatial indexes are generally categorized into three classes.
The first category is \emph{space partitioning-based indexes}, e.g., k-d trees~\cite{DBLP:journals/cacm/Bentley75}, Quad-trees~\cite{DBLP:journals/acta/FinkelB74}, and Grid Files~\cite{DBLP:journals/tods/NievergeltHS84}, which recursively split the data space into sub-regions and then index each sub-region hierarchically.
%
%
The second category is \emph{data partitioning-based indexes}, including R-tree~\cite{DBLP:conf/sigmod/Guttman84} and its variants~\cite{DBLP:conf/vldb/SellisRF87,DBLP:conf/sigmod/BeckmannKSS90,DBLP:conf/vldb/KamelF94,DBLP:conf/sigmod/ArgeBHY04,DBLP:conf/sigmod/BeckmannS09}, which recursively divide the dataset into subsets and then index each subset hierarchically.
The differences between different R-tree variants lie in how they evaluate the goodness of data partitioning and the algorithms for index construction based on the partitioning scheme.
%
The third category is \emph{data transformation-based indexes}, which transform multi-dimensional data into one dimension and then utilize a one-dimensional index, e.g., B-trees. 
%
\emph{Space-filling curves} (SFCs) are the typical transformation method.
Such data transformations act as a hybrid between data and space partitioning methods where the space partitioning nature of SFC is utilized along with data partitioning methods of the one-dimension index.
The \zindex~\cite{DBLP:conf/wwca/Bayer97,ramakrishnan2003database} is a typical index of this kind, based on the \zcurve for data transformation.
The index construction for \zindex most commonly proceeds by sorting all the points using the SFC sort-order, packing them into leaves, and then building the index bottom-up level-by-level.
Some mechanisms that utilize properties of the SFCs have been used to improve range query performance in such indexes: like the \emph{BIGMIN} algorithm associated with Z-order curves~\cite{tropf1981multidimensional} or the \emph{calculate\_next\_match} function for Hilbert curves~\cite{DBLP:conf/bncod/LawderK00}.
However, none of the above spatial indexes can be adapted to data or query patterns, and thus they often suffer from inferior query performance compared to their learned counterparts.



\spara{Learned Indexes.}
\citet{DBLP:conf/sigmod/KraskaBCDP18} first proposed the \emph{learned indexes}, which utilize machine learning models to enhance or replace traditional indexes for data access in databases.
The abstraction used in~\cite{DBLP:conf/sigmod/KraskaBCDP18} to motivate such methods was that an index is essentially a structure that predicts the location of an item in a sorted array, or a structure that predicts the cumulative density function of the underlying data.
They propose the recursive model index (RMI), which consists of a hierarchy of regression models for capturing the relationship between sorted keys and their ranks in the dataset.
According to \cite{DBLP:conf/sigmod/KraskaBCDP18}, the benefits of learned indexes lie in smaller index sizes and lower query latency. 
After this seminal work, many learned indexes were proposed in the last four years, such as PGM-Index~\cite{DBLP:journals/pvldb/FerraginaV20}, ALEX~\cite{DBLP:conf/sigmod/DingMYWDLZCGKLK20}, RadixSpline~\cite{10.1145/3401071.3401659}, and Shift-Table~\cite{DBLP:conf/edbt/0001H21}.
They outperformed RMI by using simpler linear spline models and supporting efficient index updates.
Nevertheless, all the learned indexes mentioned above are specific to one-dimensional data and cannot be directly used for spatial query processing.

\spara{Learned Spatial Indexes.} 
The most relevant studies to our problem are the ones on learned spatial indexes \cite{DBLP:conf/mdm/WangFX019,DBLP:journals/pvldb/QiLJK20,DBLP:conf/sigmod/Li0ZY020, DBLP:journals/pvldb/DingNAK20,DBLP:conf/sigmod/YangCWGLMLKA20,DBLP:conf/sigmod/NathanDAK20,DBLP:conf/ssd/ZhangRLZ21,DBLP:journals/pacmmod/GuFCL0W23,DBLP:journals/pvldb/MotiSP22,DBLP:conf/icde/DongCLLFZ22,DBLP:conf/mdm/Al-MamunHWA22}.
The learned spatial indexes can be categorized into two categories.
The first category is \emph{learned search structure-based} indexes, where the methods use learned models to improve search performance on default data layouts.
\citet{DBLP:conf/mdm/WangFX019} proposed the ZM-index, an extension of RMI~\cite{DBLP:conf/sigmod/KraskaBCDP18} using a linear layout based on Z-order values for spatial data points.
\citet{DBLP:journals/pvldb/QiLJK20} proposed a recursive spatial model index (\rsmi) to build an index utilizing the Z-order values and using neural networks to infer data partitioning.
The intuition behind RSMI is that the application of neural networks to learn fine-grained Z-order mappings would result in partitions with better locality.
While ZM and RSMI seem conceptually similar to \waz, they have three main differences from \waz.
First, ZM and RSMI utilize Z-ordering in the rank space of the data, whereas \waz avoids the rank space projection by operating in the original data space.
Second, ZM and RSMI use default Z-ordering for the data layout and a hierarchy of learned models for query processing.
However, \waz optimizes the data layout using learned models only in the training phase but does not use them to process queries.
Third, both ZM and RSMI (as with other learned search structure-based indexes) are workload-agnostic, while \waz is workload-aware.
\citet{DBLP:conf/sigmod/Li0ZY020} proposed LISA, a spatial index that divides the data space into a grid-like layout numbered by a partially monotonic function and learns functions to map data points to a grid cell and subsequently its respective page.
\citet{DBLP:conf/ssd/ZhangRLZ21} proposed SPRIG, a learned spatial index that uses the spatial interpolation function as a learned model to efficiently filter the cells in the grid file.
\citet{DBLP:conf/mdm/Al-MamunHWA22} proposed a method to improve the query performance of R-trees by using multi-label classification models to identify the required leaf nodes for a query incurring high tree-traversal cost.

The second category is \emph{layout optimization-based} indexes, where data layout is optimized using ML models to improve query performance.
\citet{DBLP:conf/sigmod/NathanDAK20} proposed a learned multi-dimensional index called \flood based on the Grid File index \cite{DBLP:journals/tods/NievergeltHS84} using machine-aware optimizations for grid layouts.
\flood is constructed by training a performance prediction model based on randomly sampled layouts, and then using the prediction model to select the best layout from a set of candidate grid layouts.
The main limitation is that \flood is optimized for an average query and performs poorly on skewed query workloads.
This limitation was addressed in \textsc{Tsunami}~\cite{DBLP:journals/pvldb/DingNAK20}, a learned multi-dimensional index that partitions the data space based on statistical tests of query distribution and constructs \flood indexes over each partition.
The statistical tests for query distribution changes are based on the shapes and sizes of queries and hence do not apply to our setting in this paper.
\citet{DBLP:conf/sigmod/YangCWGLMLKA20} proposed an index-like structure called Qd-tree to optimize multi-dimensional data layouts using deep reinforcement learning (RL).
\citet{DBLP:journals/pacmmod/GuFCL0W23} proposed a technique to construct R-trees by training RL-methods to perform ChooseSubtree and Split operations in required for sequential index construction. 
\citet{DBLP:conf/icde/DongCLLFZ22} proposed an R-tree variant that utilizes ML models to partition the dataset into homogeneous zones.
%


\section{The Base \zindex}
\label{sec:basevariant}

In this section, we provide the basic background on \zindex. 
A \zindex is a space-partitioning index that divides the data space hierarchically.
At each level, the corresponding data space is partitioned into four cells along the median of data points, and the four child cells are ordered in a `Z'-pattern.
The complete \zindex thus enforces an ordering on cells of the leaf nodes, hence imposing a partial-sort order among the data points. 

\begin{figure}[t]
  \centering

\input{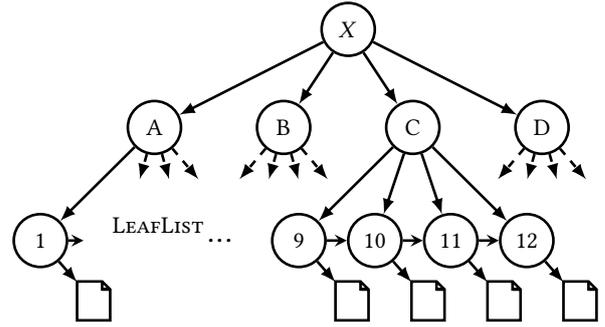}

\begin{tikzpicture}[
    roundnodes/.style = {circle,draw,minimum size = 0.7cm,line width=1pt},
    dummy/.style = {draw=none},
    empty/.style = {sibling distance = 0.45cm, yshift=0.7cm,dashed},
    secondlayer/.style = {sibling distance = 1cm,level distance =1.5cm},
    pagelayer/.style = {level distance =1cm},
    pagenodes/.style = {foldedpage,draw,minimum width=0.4cm,minimum height=0.48cm,line width=1pt,xshift=0.7cm,,yshift=0.2cm},
    edge from parent/.style={draw,-latex,line width=1pt}
]
    \node[roundnodes] {\cellX} %
        child[sibling distance = 1.7cm, yshift=0.2cm] {node[roundnodes] {\A}
            child[secondlayer] {node[roundnodes](firstleaf) {1}
                child[pagelayer] {node[pagenodes](firstpage){}}
            }
            child[empty]  {node[dummy]{}}
            child[empty]  {node[dummy]{}}
            child[empty]  {node[dummy]{}}
            }
        child[sibling distance = 1.7cm, yshift=0.2cm] {node[roundnodes] {\B}
            child[empty]  {node[dummy]{}}
            child[empty]  {node[dummy]{}}
            child[empty]  {node[dummy]{}}
            child[empty]  {node[dummy]{}}
        }
        child[sibling distance = 1.7cm, yshift=0.2cm] {node[roundnodes] {\CC}
            child[secondlayer] {node[roundnodes](1) {9}
                child[pagelayer] {node[pagenodes](0_page){}}
            }
            child[secondlayer] {node[roundnodes](2) {10}
                child[pagelayer] {node[pagenodes](1_page){}}
            }
            child[secondlayer] {node[roundnodes](3) {11}
                child[pagelayer] {node[pagenodes](2_page){}}
            }
            child[secondlayer] {node[roundnodes](4) {12}
                child[pagelayer] {node[pagenodes](3_page){}}
            }
        }
        child[sibling distance = 1.7cm, yshift=0.2cm] {node[roundnodes] {\D}
            child[empty]  {node[dummy]{}}
            child[empty]  {node[dummy]{}}
            child[empty]  {node[dummy]{}}
            child[empty]  {node[dummy]{}}
        };

    \draw[-stealth,line width=1pt] (1) -> (2);
    \draw[-stealth,line width=1pt] (2) -> (3);
    \draw[-stealth,line width=1pt] (3) -> (4);

    \node at ($(firstleaf)!.7!(1)$) {\textbf{\ldots}};
    \node at ($(firstleaf)!.2!(1)$) (secondleaf) {};
    \draw[-stealth,line width=1pt] (firstleaf) -> (secondleaf);

    \draw (-2.5,-2.6) node[]{\leaflist};

\end{tikzpicture}
  \caption{Illustration of the quaternary tree structure of a \zindex. Each leaf node holds a pointer to its subsequent leaf node as well as a pointer to the page containing its corresponding data points.}
  \label{fig:structure}
\end{figure}

Two elements define a \zindex. 
The first element is a \emph{hierarchical partitioning}  of the data space into cells, with each cell partitioned into four child cells down to a predetermined granularity of leaf-cell size \leafsize.
Corresponding split points \hierPart determines the hierarchical partitioning: for the base \zindex variant, these occur at the median of the \xx- and \yy-coordinates of all data points.
The second element is the \emph{ordering} \optorderValue of the cells at each level.
The four child cells of each parent cell of the base \zindex variant are ordered according to the ``\abcd'' pattern, whereby higher-level cells have higher ordering importance, as shown in Figure~\ref{fig:example:default}.
The curve order follows the higher-level ordering to place all leaf cells within cell \B after all the leaf cells within cell \A.
The relative order of the leaf cells in \A (or any other higher-level cell) is then determined by the second level of partitioning and ordering.
Therefore, a \zindex instance \zindexinstance is defined as a set of partition points $\hierPart=\partitionpoint$ and orderings \optorderValue associated with each of the internal nodes of a quaternary tree, as shown in Figure~\ref{fig:structure}.
Each \emph{internal node} stores the coordinates of the split point, according to which the corresponding cell is partitioned into its child cells and the ordering of the child cells.
During tree traversal, say for a point query \querypoint, we traverse down the tree by identifying the relevant child node at each internal node.
The relevant child node is computed by first comparing the query point \querypoint with the partition point \hierPart of the current node to identify whether the \querypoint resides in quadrants \A, \B, \CC, or \D (Lines~\ref{algo:treetraversal:subcell_start}--\ref{algo:treetraversal:subcell_end} in Algorithm~\ref{algo:treetraversal}) and picking the appropriate pointer based on the ordering \optorderValue used (Lines~\ref{algo:treetraversal:childid_start}--\ref{algo:treetraversal:childid_end}).

\begin{algorithm}[t]
  \small
  \caption{\treetraversal}
  \label{algo:treetraversal}
  \SetKwInOut{inputargs}{Input}
  \SetKwInOut{result}{Output}
  \SetKwFunction{FMain}{\treetraversal}
  \inputargs{\zindex \zindexinstance, node \cellX in \zindexinstance, point query \querypoint}
  \result{Boolean \bool whether \querypoint exists}

  \SetKwProg{Fn}{Function}{:}{}
  \uIf{\cellX is a leaf node}{
    \Return{$\cellX \sdot page$}
  }
  \Else{
    $bit_{x} = \querypoint \sdot \xval > \cellX \sdot \xval $ \label{algo:treetraversal:subcell_start}\\
    $bit_{y} = \querypoint \sdot \yval > \cellX \sdot \yval $ \label{algo:treetraversal:subcell_end}\\
    \uIf{$\cellX \sdot \optorderValue ==$ ``\textnormal{\abcd}''}{\label{algo:treetraversal:childid_start}
      $c_{id} = 2 bit_{y} + bit_{x}$  
    }
    \Else{
      $c_{id} = 2 bit_{x} + bit_{y}$ \label{algo:treetraversal:childid_end}
    }
    \Return{\textnormal{\treetraversal}($\cellX \sdot child\left[c_{id}\right]$)}
  }
\end{algorithm}

\begin{algorithm}[t]
  \small
  \caption{\rangequery}
  \label{algo:rangequery}
  \SetKwInOut{inputargs}{Input}
  \SetKwInOut{result}{Output}
  \inputargs{\zindex \zindexinstance, range query \query}
  \result{Set of points \queryresult in range \query}
  Page $\low = \treetraversal\left(\bottomleft\left(\query\right)\right)$ \label{algo:rangequery:bottomleft}\\
  Page $\high = \treetraversal\left(\topright\left(\query\right)\right)$ \label{algo:rangequery:topright}\\
  Initialize \queryresult = $\emptyset$ and \page = \low \label{algo:rangequery:scanstart}\\
  \While{$\page \leq \high$}{
    \overlapresult = \pagequeryoverlap(\page, \query) \label{algo:rangequery:pagequeryoverlap}\\
    \If{overlap $==$ true}{
      \queryresult $\gets$ \queryresult $\cup $ \filter(\page) 
    }
    \page = \nextpage(\page, \overlapresult) \label{algo:rangequery:nextpage} \\
  } 
  \Return{\queryresult}
\end{algorithm}

The \emph{leaf nodes} of a \zindex contain a bounding rectangle (\mbr) for the area spanned by the leaf and a pointer to a page with at most \leafsize elements.  
Each leaf node also contains a pointer to the next leaf node defined by the sort order, creating a linked list structure at the leaf layer of the tree (\leaflist).
Note that \zindex is assumed to be \emph{clustered}, with data points corresponding to consecutive leaf nodes stored in consecutive pages.
We consider the data points within a page to be stored in random order.
%

Crucially, note that the \abcd ordering guarantees \emph{monotonicity} for points that fall within different leaf cells: if point \aa in page \cellX is dominated by point \bb in page $ \cellY \neq \cellX $, then leaf \cellX will appear earlier in the \leaflist than \cellY.
Here, point \aa is dominated by point \bb if $ \aa.\xval \leq \bb.\xval $ and $ \aa.\yval \leq \bb.\yval $, and at least one coordinate of \aa is strictly smaller than that of \bb.
This monotonicity property is utilized in a \zindex for a specific range query processing mechanism.
A range query \query, defined by two points $\bottomleft(\query)$ and $\topright(\query)$ (see Figure~\ref{fig:example:defquery}), is processed by \zindex in two phases.
First, the leaf nodes enclosing two query extremes, $\bottomleft(\query)$ and $\topright(\query)$, within the \zindex are identified (Lines~\ref{algo:rangequery:bottomleft}--\ref{algo:rangequery:topright} of Algorithm~\ref{algo:rangequery}) in the \emph{index lookup phase}.
Let us refer to these leaf nodes as \low and \high.
These represent the first and last possible leaf nodes in the \leaflist that may overlap with \query.
Second, in the \emph{scanning phase}, we check the leaf nodes within the range $[\low: \high]$ for overlap with \query based on their \mbr, and pages of overlapping nodes are scanned to filter the query results (Lines~\ref{algo:rangequery:scanstart}--\ref{algo:rangequery:nextpage} of Algorithm~\ref{algo:rangequery}).
%
%


\section{The \waz Index}
\label{sec:wazi_index}

This section presents our method to build a workload-aware variant of the \zindex, i.e., \waz.
Within a \zindex, the scanning phase completely dominates the query latency.
We aim to minimize the number of points accessed during the scanning phase.
Towards this end, we present a cost formulation for the number of points accessed during range query processing (Section~\ref{sec:wazi_index:adapt}) and an approach to building \waz based on minimizing the cost function (Section~\ref{sec:wazi_index:indexconst}).

\subsection{Adaptive Partitioning and Ordering}
\label{sec:wazi_index:adapt}
The two defining elements of a \zindex, partitioning and ordering, are computed using fixed heuristics for the base variant.
By contrast, we propose a generalized variant \waz of the \zindex, for which the partitioning and ordering can vary for each node, as illustrated in Figure~\ref{fig:example:alt2levels}.
First, whereas for the base variant, the split points are predetermined to be placed at the median of the data along the \xval and \yval axes, for \waz, we propose that the split points can be placed anywhere within the data range, allowing for more flexible data partitioning.
Second, whereas for the base variant, the ordering of the child cells is predetermined to follow the ``\abcd'' pattern for every parent cell, for \waz, we propose the order is allowed to be either ``\abcd'' or ``\acbd'', as both orderings preserve monotonicity.
Our aim in making these changes in \waz is for the index to be adaptive to the given data and anticipated range queries.
For example, a \zindex with the alternative partitioning and ordering shown in Figure~\ref{fig:example:altsplitquery} would be better adapted to the range query \query shown therein than the one shown in Figure~\ref{fig:example:defquery}, as it would retrieve fewer data points in query processing.
Specifically, for a given dataset \ddistr and a set of range queries \qdistr, we are interested in building an instance of the generalized \zindex to minimize a corresponding \emph{retrieval cost}.
The retrieval cost for a range query is measured by the number of data points compared against a query box \query in the filtering phase.
In practice, \qdistr can be obtained from historical logs of range queries, as ``representatives'' for the application at hand, or in general, as anticipated range queries for which the index should be optimized.
%

\subsection{Retrieval Cost}
Here we present the objective function we aim to optimize when building a \zindex, as described above.
For a given \zindex and query \query, let $\iden{\cellX}{\cellY}$ be the function to indicate whether \query has its bottom-left vertex in \cellX and its top-right vertex in \cellY.
As a simple illustration case, let us consider a single-level \zindex with cells ordered according to the ``\abcd'' order, and the splits occur as in Figure~\ref{fig:example:defquery}.
Following the range query processing discussed earlier, the retrieval cost for the chosen split and ordering equals
\begin{align}
  \normcost{\cellX} & (\query \,|\, \xval, \yval; \abcd) = \iden{\A}{\D} (\nn{\A} + \nn{\B} +\nn{\CC} + \nn{\D} ) + \nonumber \\
  &  \iden{\A}{\CC}(\nn{\A} + \pagefactor\nn{\B} + \nn{\CC}) + \iden{\B}{\D} (\nn{\B} + \pagefactor\nn{\CC} + \nn{\D}) + \nonumber \\
  & \iden{\A}{\B}(\nn{\A} + \nn{\B}) + \iden{\CC}{\D} (\nn{\CC} + \nn{\D}) +  \nonumber \\
  & \iden{\A}{\A}\nn{\A} + \iden{\B}{\B}\nn{\B} + \iden{\CC}{\CC}\nn{\CC} + \iden{\D}{\D}\nn{\D}, \label{eq:singlelevelcostperquery}
\end{align}
where \nn{\cellX} denotes the number of data points in each cell.
Notice that \nn{\cellX} and $\iden{\cellX}{\cellY}$ depend on the split location $(\xval, \yval)$; however, we omit the dependency from our notation for simplicity.
To see why the formula holds, note that when \cellrange{\query}{\A}{\B}, the \zindex retrieves all points only from cells \A and \B, as no other cells fall between \A and \B in the ``\abcd'' ordering of cells.
However, when \cellrange{{\query}}{\A}{\CC} (as in Figure~\ref{fig:example:defquery}), in addition to filtering points from \A and \CC, the \zindex also compares \mbr and skips over the non-overlapping leaf node \B as it falls between \A and \CC in the ``\abcd'' ordering.
The impact of skipping over leaf cells is represented in our cost as a fraction $\pagefactor<1$ of the number of points.
The rest of the cases follow similarly.
Notice that if the ordering of the cells is ``\acbd'' instead, the cost formula will differ from Eq.~\ref{eq:singlelevelcostperquery}:
\begin{align}
  \normcost{\cellX} & (\query\,|\,\xval, \yval; \acbd) = \iden{\A}{\D} (\nn{\A} + \nn{\B} +\nn{\CC} + \nn{\D} ) + \nonumber \\
  & \iden{\A}{\B}(\nn{\A} + \pagefactor\nn{\B} + \nn{\CC}) + \iden{\CC}{\D} (\nn{\B} + \pagefactor\nn{\CC} + \nn{\D}) + \nonumber \\
  & \iden{\A}{\CC}(\nn{\A} + \nn{\CC}) + \iden{\B}{\D} (\nn{\B} + \nn{\D}) + \nonumber \\
  & \iden{\A}{\A}\nn{\A} + \iden{\B}{\B}\nn{\B} + \iden{\CC}{\CC}\nn{\CC} + \iden{\D}{\D}\nn{\D} \label{eq:altsinglelevelcostperquery}
\end{align}
%

More generally, when the \zindex consists of more than one level of partitions, the retrieval cost is defined recursively, as the \zindex structure of second-level partitions (for each of \A, \B, \CC, and \D) affects the total cost of retrieval. 
Defining the retrieval cost of a \zindex with two levels would involve recursively substituting Eq.~\ref{eq:singlelevelcostperquery} or~\ref{eq:altsinglelevelcostperquery} for each term of the form \iden{\cellX}{\cellX}\nn{\cellX} based on the ordering $\orderValue_\cellX$ under consideration: 
\begin{align}
  \normcost{\cellX} & (\query\,|\,\xval, \yval; \acbd) = \iden{\A}{\D} (\nn{\A} + \nn{\B} +\nn{\CC} + \nn{\D} ) + \nonumber \\
  & \iden{\A}{\B}(\nn{\A} + \pagefactor\nn{\B} + \nn{\CC}) + \iden{\CC}{\D} (\nn{\B} + \pagefactor\nn{\CC} + \nn{\D}) + \nonumber \\
  & \iden{\A}{\CC}(\nn{\A} + \nn{\CC}) + \iden{\B}{\D} (\nn{\B} + \nn{\D}) + \nonumber \\
  & \normcost{\A}  (\query\,|\,\xval, \yval; \orderValue_{\A}) + \normcost{\B}  (\query\,|\,\xval, \yval; \orderValue_{\B}) + \nonumber \\
  & \normcost{\CC}  (\query\,|\,\xval, \yval; \orderValue_{\CC}) + \normcost{\D}  (\query\,|\,\xval, \yval; \orderValue_{\D}) \label{eq:recursive_cost}
\end{align}

%
For a set of queries \qdistr and a \zindex \zindexinstance with cell \cellX, the total cost of all queries is aggregated to form the full cost \totalcost{}.
Specifically, if the ordering of \cellX is ``\abcd'', the cost function is
\begin{align}
   \totalcost{\cellX} & (\qdistr\,|\,\xval, \yval; \abcd) = \sum_{\query \in \qdistr} \normcost{\cellX}(\query\,|\,\xval, \yval; \abcd) \nonumber \\
   = &\ \  \qq{\A}{\D} (\nn{\A} + \nn{\B} +\nn{\CC} + \nn{\D} ) + \nonumber \\
   & \qq{\A}{\CC}(\nn{\A} + \pagefactor\nn{\B} + \nn{\CC}) + \qq{\B}{\D} (\nn{\B} + \pagefactor\nn{\CC} + \nn{\D}) + \nonumber \\
   & \qq{\A}{\B}(\nn{\A} + \nn{\B}) + \qq{\CC}{\D} (\nn{\CC} + \nn{\D}) +  \nonumber \\
   & \totalcost{\A}  (\query\,|\,\xval, \yval; \orderValue_{\A}) + \totalcost{\B}  (\query\,|\,\xval, \yval; \orderValue_{\B}) + \nonumber \\
   & \totalcost{\CC}  (\query\,|\,\xval, \yval; \orderValue_{\CC}) + \totalcost{\D}  (\query\,|\,\xval, \yval; \orderValue_{\D}) \label{eq:recursive_totalcost},
\end{align}
with the terms $\qq{\cellX}{\cellY} = \sum_{\query \in \qdistr} \iden{\cellX}{\cellY}$.
A similar equation is also present for the alternative ordering of ``\acbd''.
%



\subsection{Index Construction}
\label{sec:wazi_index:indexconst}

The formulation of retrieval cost in Eq.~\ref{eq:recursive_totalcost} leads to a cost function that exhibits an optimal substructure, where optimal configuration and the associated ordering for all possible child cell combinations should be known before one can compute the optimal configuration at a given node.
Finding the optimal solution under this cost formulation using dynamic programming has a complexity of $\bigo{\numdata^4}$.
The complexity follows because, for \numdata points in two dimensions, there are $\numdata^4$ rectangles enclosing unique subsets of points.
Hence, the state space for dynamic programming is at most $\numdata^4$.
Obviously, such an approach is infeasible even for moderately sized datasets.

%

%
Instead, we adopt a \greedy algorithm for index construction.
The \greedy algorithm simplifies Eq.~\ref{eq:recursive_totalcost} by formulating the cost for lower levels as \qq{\cellX}{\cellX}\nn{\cellX}.
This simplification uses an upper bound on the possible retrieval cost of each sub-partition in place of the recursive cost definition of Eq.~\ref{eq:recursive_totalcost}.
Following this intuition also yields an approach that allows for optimization steps to be performed greedily for each layer hierarchically.
Intuitively, the greedy algorithm proceeds at one level at a time, from top to bottom, selecting the partition point and ordering using the alternative cost function:
\begin{align}
  \totalcost{\cellX} & (\qdistr\,|\,\xval, \yval; \abcd) = \sum_{\query \in \qdistr} \normcost{\cellX}(\query\,|\,\xval, \yval; \abcd) \nonumber \\
  = &\ \ \qq{\A}{\D} (\nn{\A} + \nn{\B} +\nn{\CC} + \nn{\D} ) + \nonumber \\
  & \qq{\A}{\CC}(\nn{\A} + \pagefactor\nn{\B} + \nn{\CC}) + \qq{\B}{\D} (\nn{\B} + \pagefactor\nn{\CC} + \nn{\D}) + \nonumber \\
  & \qq{\A}{\B}(\nn{\A} + \nn{\B}) + \qq{\CC}{\D} (\nn{\CC} + \nn{\D}) +  \nonumber \\
  & \qq{\A}{\A}\nn{\A} + \qq{\B}{\B}\nn{\B} + \qq{\CC}{\CC}\nn{\CC} + \qq{\D}{\D}\nn{\D} \label{eq:greedy_totalcost}
\end{align}

\begin{algorithm}[t]
  \small
  \caption{\greedy}
  \label{algo:greedy}

  \SetKwInOut{inputargs}{Input}
  \SetKwInOut{result}{Output}
  \SetKwFunction{SolveCell}{SolveCell}
  \SetKwFunction{Greedy}{Greedy}
  \SetKwFunction{uniformsearch}{UniformSearchCandidates}
  \SetKwFunction{uniformsampl}{UniformSample}
  \SetKwFunction{split}{Split}

  \inputargs{\zindex \zindexinstance, node \cellX in \zindexinstance, workload \qdistr, data \ddistr}

  \lIf{$\nn{\cellX} < \leafsize$}{
    \Return
  }
  \tcc {Draw \numsamples candidate split points}
  $XY$\,:= \uniformsampl(\cellX,\numsamples) \\
  \tcc {Select split and ordering to minimize Eq.~\ref{eq:greedy_totalcost}}
  (\xopt, \yopt, \optorderValue) = $\arg\min_{(\xval,\yval) \in XY; \optorderValue\in\{\abcd, \acbd\} } \totalcost{}(\qdistr\,|\,\xval, \yval; \orderValue)$ \label{algo:greedy:cost}\\
  \tcc {Define cells w.r.t.~split point}
  Cells $\A,\B,\CC,\D$\,:= \split{\cellX, $(\xopt, \yopt)$} \\
  Add cells $\A,\B,\CC,\D$ and ordering \optorderValue to \zindexinstance\\
  \tcc {Apply Greedy to child cells}
  \ForEach{Cell $\cellY \in \{\A, \B, \CC, \D\}$}
  {
    \greedy(\zindexinstance, \cellY, \qdistr, \ddistr; \SolveCell)
  }
\end{algorithm}
The pseudocode for index construction is presented in Algorithm \ref{algo:greedy}.
The steps for index construction are described at three levels of detail.
First, our algorithm proceeds greedily, determining the partitioning and ordering of the cells within the same level, one level at a time, from top (root) to bottom (leaf).
Therefore, every time a cell \cellX needs to be split and its children ordered, we use the configuration that minimizes the objective \totalcost{} (Eq.~\ref{eq:greedy_totalcost}).
Second, we choose the partitioning and ordering minimizing the objective \totalcost{} for a given cell \cellX by uniformly sampling the candidate split points and selecting the one that minimizes our objective.
More specifically, for each cell that it considers for splitting, \greedy samples \numsamples candidate split points uniformly at random from the region covered by the cell, evaluates the objective \totalcost{} for each candidate split point for both possible orderings, and returns the split-point and ordering with the minimum \totalcost{}.
We choose sampling-based optimization over more complicated optimizers (like Bayesian optimization) to avoid the high computation overhead incurred by such optimizers.
More importantly, we observed that each iteration of an optimizer incurs computational cost several magnitudes higher than computing the objective function. 
In our experiments, we observe no performance improvement over sampling-based optimization to justify the added computational cost of said optimizers.
Third, we approximate the exact data distribution \ddistr and range query distribution \qdistr with approximate distributions by ML models (i.e., ``learned'' approaches).
These approximate distributions allow for efficient estimations of the number of data points (and queries) that fall within each of the child cells resulting from a candidate partitioning of a given cell, as needed to compute the objective \totalcost{}.
We used Random Forest Density Estimation models (RFDE) \cite{DBLP:conf/icml/WenH22} for our density estimation.
Specifically, we construct a forest of k-d trees constructed with randomized split dimensions at each node.
Each node stores the number of data points (\emph{cardinality}) contained within the region. 
Density estimation in an RFDE model is performed as a tree traversal, collecting the cardinality information from nodes overlapping the density estimation query.


\begin{figure*}[t]
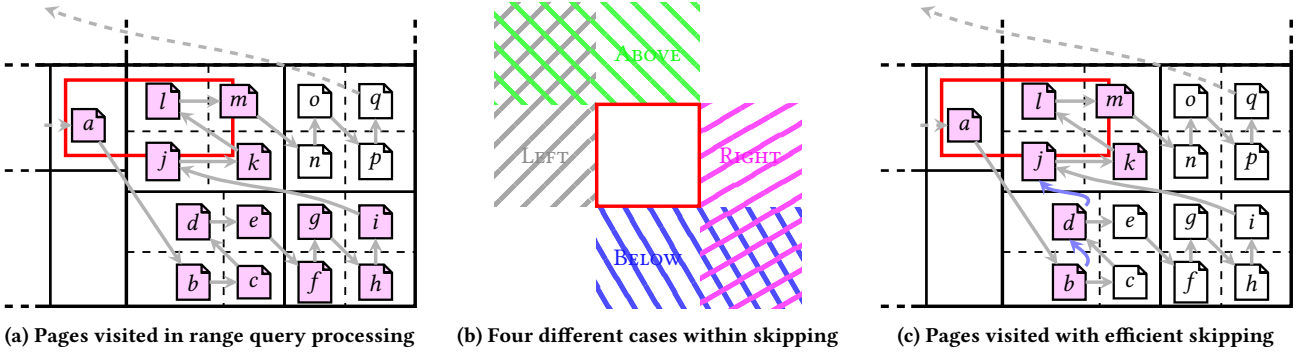

    \centering
    \begin{subfigure}{0.33\textwidth}
        \centering
        \tikz[line width=0.3mm,scale=2.0]{


\input{./figures/tikz_figures/pageshape.tex}


\input{./figures/tikz_figures/pageshape.tex}

\tikzstyle{pagenodes} = [foldedpage,draw,minimum width=0.4cm,minimum height=0.4cm]
\tikzset{edge from parent/.style={draw,-latex}}
\tikzstyle{pagezorder} = [line width=1.3pt, gray!60]

\coordinate (TL) at (-0.1,-0.5);
\coordinate (BR) at (1,-1);

\newcommand \upperlimit {-0.1} ;
\newcommand \lowerlimit {-2} ;

\newcommand \leftlimit {-0.3} ;
\newcommand \rightlimit {2.2} ;

\newcommand \levelonex {0.3};
\newcommand \leveloney {-0.4};

\newcommand \leveltwox {1.34};
\newcommand \leveltwoy {-1.24};

\newcommand \dashedsize {0.2};

\draw [line width=1.3pt,dashed] (\rightlimit,\upperlimit) -- (\rightlimit,\upperlimit-\dashedsize);
\draw [line width=1.3pt] (\rightlimit,\upperlimit-\dashedsize) -- (\rightlimit,\lowerlimit) -- (\leftlimit,\lowerlimit);
\draw [line width=1.3pt,dashed] (\leftlimit-\dashedsize,\lowerlimit) -- (\leftlimit,\lowerlimit);

\draw [line width=1.3pt,dashed] (\levelonex,\upperlimit)  -- (\levelonex,\upperlimit-\dashedsize);
\draw [line width=1.3pt] (\levelonex,\upperlimit-\dashedsize)  -- (\levelonex,\lowerlimit);

\draw [line width=1.3pt,dashed] (\leftlimit-\dashedsize,\leveloney) -- (\rightlimit,\leveloney);
\draw [line width=1.3pt] (\leftlimit,\leveloney) -- (\rightlimit,\leveloney);

\draw[line width=1pt,dashed] (\leftlimit-\dashedsize,-1.1) -- (\leftlimit,-1.1);
\draw[line width=1pt] (\leftlimit,-1.1) -- (\levelonex,-1.1);
\draw[line width=1pt] (\leveltwox,\leveloney) -- (\leveltwox , \lowerlimit);

\draw[line width=1pt] (-0.2,\leveloney) -- (-0.2,\lowerlimit);

\draw[line width=1pt] (\levelonex,\leveltwoy) -- (\rightlimit , \leveltwoy);
\draw[line width=1pt] (\leveltwox,\leveloney) -- (\leveltwox , \lowerlimit);

\draw[line width=0.7pt,dashed] (\levelonex, \leveltwoy-0.4) -- (\leveltwox , \leveltwoy-0.4);
\draw[line width=0.7pt,dashed] (\leveltwox - 0.4 ,\leveltwoy) -- (\leveltwox - 0.4,\lowerlimit);

\draw[line width=0.7pt,dashed] (\leveltwox, \leveltwoy-0.4) -- (\rightlimit , \leveltwoy-0.4);
\draw[line width=0.7pt,dashed] (\leveltwox+0.4,\leveltwoy) -- (\leveltwox+0.4,\lowerlimit);

\draw[line width=0.7pt,dashed] (\levelonex,\leveltwoy + 0.4) -- (\leveltwox ,\leveltwoy + 0.4);
\draw[line width=0.7pt,dashed] (\leveltwox-0.5,\leveloney) -- (\leveltwox-0.5,\leveltwoy);

\draw[line width=0.7pt,dashed] (\leveltwox, \leveltwoy+0.4) -- (\rightlimit , \leveltwoy+0.4);
\draw[line width=0.7pt,dashed] (\leveltwox+0.4,\leveloney) -- (\leveltwox+0.4,\leveltwoy);

\draw[red,line width=1.3pt] (TL) rectangle (BR);

\draw (\leftlimit,-0.8) node(minus_a){}; 
\draw (0.05,-0.8) node[pagenodes,fill=magenta!20](page_a){$a$}; 

\draw (\leveltwox-0.6,\leveltwoy-0.6) node[pagenodes,fill=magenta!20](page_b){$b$};
\draw (\leveltwox-0.2,\leveltwoy-0.6) node[pagenodes,fill=magenta!20](page_c){$c$};
\draw (\leveltwox-0.6,\leveltwoy-0.2) node[pagenodes,fill=magenta!20](page_d){$d$};
\draw (\leveltwox-0.2,\leveltwoy-0.2) node[pagenodes,fill=magenta!20](page_e){$e$};

\draw (\leveltwox+0.2,\leveltwoy-0.6) node[pagenodes,fill=magenta!20](page_f){$f$};
\draw (\leveltwox+0.2,\leveltwoy-0.2) node[pagenodes,fill=magenta!20](page_g){$g$};
\draw (\leveltwox+0.6,\leveltwoy-0.6) node[pagenodes,fill=magenta!20](page_h){$h$};
\draw (\leveltwox+0.6,\leveltwoy-0.2) node[pagenodes,fill=magenta!20](page_i){$i$};

\draw (\leveltwox-0.8,\leveltwoy+0.2) node[pagenodes,fill=magenta!20](page_j){$j$};
\draw (\leveltwox-0.2,\leveltwoy+0.2) node[pagenodes,fill=magenta!20](page_k){$k$};
\draw (\leveltwox-0.8,\leveltwoy+0.6) node[pagenodes,fill=magenta!20](page_l){$l$};
\draw (\leveltwox-0.3,\leveltwoy+0.6) node[pagenodes,fill=magenta!20](page_m){$m$};

\draw (\leveltwox+0.2,\leveltwoy+0.2) node[pagenodes](page_n){$n$};
\draw (\leveltwox+0.2,\leveltwoy+0.6) node[pagenodes](page_o){$o$};
\draw (\leveltwox+0.6,\leveltwoy+0.2) node[pagenodes](page_p){$p$};
\draw (\leveltwox+0.6,\leveltwoy+0.6) node[pagenodes](page_q){$q$};

\draw (\leftlimit,\leveloney+0.4) node(plus_q){};

\draw[-stealth,pagezorder,dashed] (minus_a) -> (page_a);

\draw[-stealth,pagezorder] (page_a) -> (page_b);
\draw[-stealth,pagezorder] (page_b) -> (page_c);
\draw[-stealth,pagezorder] (page_c) -> (page_d);
\draw[-stealth,pagezorder] (page_d) -> (page_e);
\draw[-stealth,pagezorder] (page_e) -> (page_f);
\draw[-stealth,pagezorder] (page_f) -> (page_g);
\draw[-stealth,pagezorder] (page_g) -> (page_h);
\draw[-stealth,pagezorder] (page_h) -> (page_i);
\draw[-stealth,pagezorder] (page_i) to [out=155,in= -20] (page_j);
\draw[-stealth,pagezorder] (page_j) -> (page_k);
\draw[-stealth,pagezorder] (page_k) -> (page_l);
\draw[-stealth,pagezorder] (page_l) -> (page_m);
\draw[-stealth,pagezorder] (page_m) -> (page_n);
\draw[-stealth,pagezorder] (page_n) -> (page_o);
\draw[-stealth,pagezorder] (page_o) -> (page_p);
\draw[-stealth,pagezorder] (page_p) -> (page_q);

\draw[-stealth,pagezorder,dashed] (page_q) to [out=155,in= -20] (plus_q);



}
        \caption{Pages visited in range query processing}
        \label{fig:skippingneed}
    \end{subfigure}
    \hfill
    \begin{subfigure}{0.33\textwidth}
        \centering
        \tikz[line width=0.3mm,scale=1.8]{\input{./figures/tikz_figures/skipping_cases.tex}}
        \caption{Four different cases within skipping}
        \label{fig:skippingcases}
    \end{subfigure}
    \hfill
    \begin{subfigure}{0.33\textwidth}
        \centering
        \tikz[line width=0.3mm,scale=2.0]{


\input{./figures/tikz_figures/pageshape.tex}


\input{./figures/tikz_figures/pageshape.tex}

\tikzstyle{pagenodes} = [foldedpage,draw,minimum width=0.4cm,minimum height=0.4cm]
\tikzset{edge from parent/.style={draw,-latex}}
\tikzstyle{pagezorder} = [line width=1.3pt, gray!60]

\coordinate (TL) at (-0.1,-0.5);
\coordinate (BR) at (1,-1);

\newcommand \upperlimit {-0.1} ;
\newcommand \lowerlimit {-2} ;

\newcommand \leftlimit {-0.3} ;
\newcommand \rightlimit {2.2} ;

\newcommand \levelonex {0.3};
\newcommand \leveloney {-0.4};

\newcommand \leveltwox {1.34};
\newcommand \leveltwoy {-1.24};

\newcommand \dashedsize {0.2};

\draw [line width=1.3pt,dashed] (\rightlimit,\upperlimit) -- (\rightlimit,\upperlimit-\dashedsize);
\draw [line width=1.3pt] (\rightlimit,\upperlimit-\dashedsize) -- (\rightlimit,\lowerlimit) -- (\leftlimit,\lowerlimit);
\draw [line width=1.3pt,dashed] (\leftlimit-\dashedsize,\lowerlimit) -- (\leftlimit,\lowerlimit);

\draw [line width=1.3pt,dashed] (\levelonex,\upperlimit)  -- (\levelonex,\upperlimit-\dashedsize);
\draw [line width=1.3pt] (\levelonex,\upperlimit-\dashedsize)  -- (\levelonex,\lowerlimit);

\draw [line width=1.3pt,dashed] (\leftlimit-\dashedsize,\leveloney) -- (\rightlimit,\leveloney);
\draw [line width=1.3pt] (\leftlimit,\leveloney) -- (\rightlimit,\leveloney);

\draw[line width=1pt,dashed] (\leftlimit-\dashedsize,-1.1) -- (\leftlimit,-1.1);
\draw[line width=1pt] (\leftlimit,-1.1) -- (\levelonex,-1.1);
\draw[line width=1pt] (\leveltwox,\leveloney) -- (\leveltwox , \lowerlimit);

\draw[line width=1pt] (-0.2,\leveloney) -- (-0.2,\lowerlimit);

\draw[line width=1pt] (\levelonex,\leveltwoy) -- (\rightlimit , \leveltwoy);
\draw[line width=1pt] (\leveltwox,\leveloney) -- (\leveltwox , \lowerlimit);

\draw[line width=0.7pt,dashed] (\levelonex, \leveltwoy-0.4) -- (\leveltwox , \leveltwoy-0.4);
\draw[line width=0.7pt,dashed] (\leveltwox - 0.4 ,\leveltwoy) -- (\leveltwox - 0.4,\lowerlimit);

\draw[line width=0.7pt,dashed] (\leveltwox, \leveltwoy-0.4) -- (\rightlimit , \leveltwoy-0.4);
\draw[line width=0.7pt,dashed] (\leveltwox+0.4,\leveltwoy) -- (\leveltwox+0.4,\lowerlimit);

\draw[line width=0.7pt,dashed] (\levelonex,\leveltwoy + 0.4) -- (\leveltwox ,\leveltwoy + 0.4);
\draw[line width=0.7pt,dashed] (\leveltwox-0.5,\leveloney) -- (\leveltwox-0.5,\leveltwoy);

\draw[line width=0.7pt,dashed] (\leveltwox, \leveltwoy+0.4) -- (\rightlimit , \leveltwoy+0.4);
\draw[line width=0.7pt,dashed] (\leveltwox+0.4,\leveloney) -- (\leveltwox+0.4,\leveltwoy);

\draw[red,line width=1.3pt] (TL) rectangle (BR);

\draw (\leftlimit,-0.8) node(minus_a){}; 
\draw (0.05,-0.8) node[pagenodes,fill=magenta!20](page_a){$a$}; 

\draw (\leveltwox-0.6,\leveltwoy-0.6) node[pagenodes,fill=magenta!20](page_b){$b$};
\draw (\leveltwox-0.2,\leveltwoy-0.6) node[pagenodes](page_c){$c$};
\draw (\leveltwox-0.6,\leveltwoy-0.2) node[pagenodes,fill=magenta!20](page_d){$d$};
\draw (\leveltwox-0.2,\leveltwoy-0.2) node[pagenodes](page_e){$e$};

\draw (\leveltwox+0.2,\leveltwoy-0.6) node[pagenodes](page_f){$f$};
\draw (\leveltwox+0.2,\leveltwoy-0.2) node[pagenodes](page_g){$g$};
\draw (\leveltwox+0.6,\leveltwoy-0.6) node[pagenodes](page_h){$h$};
\draw (\leveltwox+0.6,\leveltwoy-0.2) node[pagenodes](page_i){$i$};

\draw (\leveltwox-0.8,\leveltwoy+0.2) node[pagenodes,fill=magenta!20](page_j){$j$};
\draw (\leveltwox-0.2,\leveltwoy+0.2) node[pagenodes,fill=magenta!20](page_k){$k$};
\draw (\leveltwox-0.8,\leveltwoy+0.6) node[pagenodes,fill=magenta!20](page_l){$l$};
\draw (\leveltwox-0.3,\leveltwoy+0.6) node[pagenodes,fill=magenta!20](page_m){$m$};

\draw (\leveltwox+0.2,\leveltwoy+0.2) node[pagenodes](page_n){$n$};
\draw (\leveltwox+0.2,\leveltwoy+0.6) node[pagenodes](page_o){$o$};
\draw (\leveltwox+0.6,\leveltwoy+0.2) node[pagenodes](page_p){$p$};
\draw (\leveltwox+0.6,\leveltwoy+0.6) node[pagenodes](page_q){$q$};

\draw (\leftlimit,\leveloney+0.4) node(plus_q){};

\draw[-stealth,pagezorder,dashed] (minus_a) -> (page_a);

\draw[-stealth,pagezorder] (page_a) -> (page_b);
\draw[-stealth,pagezorder] (page_b) -> (page_c);
\draw[-stealth,pagezorder] (page_c) -> (page_d);
\draw[-stealth,pagezorder] (page_d) -> (page_e);
\draw[-stealth,pagezorder] (page_e) -> (page_f);
\draw[-stealth,pagezorder] (page_f) -> (page_g);
\draw[-stealth,pagezorder] (page_g) -> (page_h);
\draw[-stealth,pagezorder] (page_h) -> (page_i);
\draw[-stealth,pagezorder] (page_i) to [out=155,in= -20] (page_j);
\draw[-stealth,pagezorder] (page_j) -> (page_k);
\draw[-stealth,pagezorder] (page_k) -> (page_l);
\draw[-stealth,pagezorder] (page_l) -> (page_m);
\draw[-stealth,pagezorder] (page_m) -> (page_n);
\draw[-stealth,pagezorder] (page_n) -> (page_o);
\draw[-stealth,pagezorder] (page_o) -> (page_p);
\draw[-stealth,pagezorder] (page_p) -> (page_q);

\draw[-stealth,pagezorder,dashed] (page_q) to [out=155,in= -20] (plus_q);

\draw[-stealth,pagezorder,blue!50] (page_b) to [out=45,in= -45] (page_d.south);

\draw[-stealth,pagezorder,blue!50] (page_d) to [out=45,in= -45] (page_j.south);


}
        \caption{Pages visited with efficient skipping}
        \label{fig:skippingexample}
    \end{subfigure}
    \caption{Illustration of skipping during range query processing; (a) Standard range query processing of range query $R$ (red) processes pages in range $[a:m]$; (b) The four different irrelevancy criteria explained.  (c) Motivating example for efficient skipping. As we process page $b$, we know that it does not overlap the query because of \qbelow. We also know that the next page in the sort-order that may satisfy the criterion is $d$. Similarly, at page $d$ we can skip ahead to page $j$, saving the computation required to process pages $e$ through $i$;}
    \label{fig:skipping}
\end{figure*}

\section{Skipping Mechanism}
\label{sec:skipping}

\begin{algorithm}[t]
    \small
    \caption{Look-ahead Pointer Construction}
    \label{algo:computelookahead}
    \SetKwInOut{inputargs}{Input}
    \SetKwFunction{InitializeLookahead}{InitializeLookahead}
    \SetKwFunction{nextpage}{next}

    \inputargs{\zindex \zindexinstance}
    \tcc{Initalize all lookahead pointer}
    \InitializeLookahead(\leaflist) \\
    \tcc{Iterate backward through leaf nodes}
    \ForEach{node \page in Reverse(\textnormal{\leaflist})}{
        \ForEach{\qcase in \textnormal{[\qbelow, \qabove, \qleft, \qright]}}{
            \tcc{Init with the next ptr of LeafList}
            \page.\qcase = \nextpage(\page)\\
            \While{\textnormal{\qcase} not improved}{
                \page.\qcase = (\page.\qcase).\qcase \label{algo:computelookahead:followpointer}
            }
        }
    }
\end{algorithm}

Processing range queries involves scanning and filtering points from an interval $[\low : \high]$ of leaf nodes within the \leaflist ordered by a \zindex.
Figure~\ref{fig:skippingneed} illustrates such an ordering (solid grey arrows) of leaf nodes.
The leaf nodes have been named $a$ through $q$ following the ordering shown in grey.
Given a range query \query (red box), the query processing algorithm will compare all leaf nodes within the interval $[a:m]$ (shaded in red) and return points from relevant leaf nodes (i.e., those that overlap with the range query).
For the example in Figure~\ref{fig:skippingneed}, the relevant leaf nodes are $a, j, k, l$, and $m$, whereas the pages $b$ through $i$ are irrelevant.
We perform a bounding box comparison of the pages before scanning points stored within them and hence data on pages $b$ through $i$ are not scanned.
But, the number of points $r$ returned by the algorithm for a range query is often comparable (or significantly smaller in case of low selectivity queries) than the total number of leaf nodes $s$ for which we compare overlap, that is, $s \gg r$.
In such cases, the redundant computation of checking overlap between bounding boxes of irrelevant leaf nodes and the range query can become a bottleneck. 
Our solution to address this issue is skipping over leaf nodes irrelevant to the query.
For the example above, Figure~\ref{fig:skippingexample} depicts two practical instances of skipping from $b$ to $d$ and $d$ to $j$ (blue arrows), reducing the number of leaf nodes visited (shaded red) from 13 to 7.
We propose our novel mechanism for efficient skipping in two parts.
First, we design a skipping mechanism that operates on any range query using \emph{look-ahead pointers}.
Second, we present the algorithm to precompute the look-ahead pointers during index construction.
Finally, we modify the retrieval cost from Eq.~\ref{eq:greedy_totalcost} to accurately incorporate the impact of skipping.
\subsection{Look-ahead Pointers}\label{subsec:efficient_skipping}
%
In standard processing (cf. Section~\ref{sec:basevariant}) for a range query \query, we check whether the bounding box of any leaf node overlaps with \query.
If so, we read and filter points stored in the page associated with that leaf node.
Otherwise, we follow the pointer to the next leaf node in the \leaflist, iteratively proceeding until a relevant page is found or we reach the end of the interval $[\low :\high]$.
As discussed earlier, this may be inefficient.
Our solution introduces four additional look-ahead pointers for each leaf node, allowing us to skip irrelevant pages accessed with the standard query processing algorithm.
The four look-ahead pointers map to the four possible criteria under which a leaf node may be irrelevant; we name them \qbelow, \qabove, \qleft, and \qright.
For instance, \qbelow indicates that the $y$-coordinate of the top-right of the leaf node \page, represented by $\topright(\page).\yval$, is lower than the $y$-coordinate of bottom-left of the query \query, $\topright(\page).\yval<\bottomleft(\query).\yval$.
Or put simply, the area covered by \page lies entirely below the range query \query.
The other criteria follow similarly.
Figure~\ref{fig:skippingcases} shows the four criteria for a candidate \query.
For example, in Figure~\ref{fig:skippingneed}, the leaf node represented by $d$ is irrelevant to the range query shown in red as it would satisfy \qbelow criteria, where $f$ satisfies both \qright and \qbelow criteria.
Note that pages within the interval $[\low:\high]$ of a query \query cannot lie in the bottom-left and top-right sections due to monotonicity constraint.
A look-ahead pointer associated with each of the four criteria points to the next possible leaf node that could be relevant, skipping over leaf nodes that are guaranteed to be deemed irrelevant due to the same criteria.
Specifically, consider a leaf node $\page_1$ whose look-ahead pointer associated with \qbelow points to $\page_2$.
$\page_2$ is the earliest leaf node in \leaflist that satisfies the conditions $\topright(\page_1).\yval < \topright(\page_2).\yval$ and $\rank(\page_1)<\rank(\page_2)$.
For example, leaf node $d$ in Figure~\ref{fig:skippingneed} is irrelevant due to the \qbelow criterion.
Consequently, nodes $[e:i]$ are guaranteed to be irrelevant to \query due to \qbelow as $\forall x\in[e:i],\topright(x).\yval \leq \topright(d).\yval$.
Therefore, we know that for any query \query that disqualifies $d$ due to \qbelow, nodes $[e:i]$ will also be deemed irrelevant.
We utilize look-ahead pointers to modify the range-query processing algorithm.
If we identify that a leaf node \page does not overlap with \query, we now follow a look-ahead pointer instead of following the next pointer.
The choice of look-ahead pointer is made by discerning the criteria under which leaf node \page was deemed irrelevant. 
If an irrelevant leaf node satisfies multiple criteria, we pick the look-ahead pointer that skips over the greatest number of nodes.

\subsection{Building Look-ahead Pointers}\label{subsec:lookhahead_pointer}
The algorithm for computing the look-ahead pointers is presented in Algorithm~\ref{algo:computelookahead}.
The look-ahead pointers are constructed in the final phase of index construction for a \zindex, where the hierarchical structure already imposes an ordering \rank on the leaf nodes.
The construction of look-ahead pointers considers leaf nodes in the reverse order of the \leaflist.
The look-ahead pointers for the last leaf node point to a dummy page signifying the end of \leaflist.
For each subsequent (i.e., earlier in \leaflist) leaf node \page, we utilize the constructed look-ahead pointers within the suffix of \leaflist to compute the look-ahead pointer. 
To construct a look-ahead pointer associated with a given criterion (say, \qbelow), we temporarily assign the corresponding look-ahead pointer $\page.\qbelow$ to the next pointer of \leaflist.
We then recursively check whether the node pointed to by $\page.\qbelow$ improves the criterion.
Improving the criterion in this case refers to having an improved value for the corresponding disqualifying comparison.
In the case of \qbelow, the improved value would be if the pointer $\page.\qbelow$ points to a page with a greater upper bound value along the $y$-coordinate; put simply, it is higher than leaf node \page.
If the check for improving criterion fails, we follow the corresponding nodes's pointer for the criterion, thus finding later nodes recursively (Line~\ref{algo:computelookahead:followpointer} of Algorithm~\ref{algo:computelookahead}).
For each leaf node, the look-ahead construction algorithm performs at most $\sqrt{|\leaflist|}$ recursion steps. 
Therefore, the complexity of Algorithm~\ref{algo:computelookahead} is $\bigo{|\leaflist|^{3/2}}$, where $|\leaflist|\approx \numdata / \leafsize$.

%
The retrieval cost formulation presented in Eq.~\ref{eq:greedy_totalcost} accounts for the skipping cost of irrelevant leaf nodes using \pagefactor.
We can now update the cost formulation to accurately reflect the retrieval cost of a given query in light of the skipping mechanism mentioned above and account for the redundant quadrant fetched for processing queries by setting the \pagefactor value to a small constant.
In our experiments, we set \pagefactor to $10^{-5}$.
The small value of \pagefactor will more accurately reflect the cost of skipping over irrelevant pages within redundant quadrants using the look-ahead pointers.
The index construction (Algorithm~\ref{algo:greedy}) remains unchanged except for the fact that one would utilize a smaller \pagefactor value for Eq.~\ref{eq:greedy_totalcost} in Line~\ref{algo:greedy:cost} of Algorithm~\ref{algo:greedy} when used along with look-ahead pointers.
%

\section{Experiments}
\label{sec:exp}

In this section, we first describe the setup of our experiments.
Then, we present the experimental results with discussions.

\subsection{Baselines}
\label{sec:exp:baselines}

We compare the learned workload-aware Z-Index \waz with the following competitors:
\begin{enumerate}[1)]
  \item Sort Tile Recursive R-tree (\str) \cite{DBLP:conf/icde/LeuteneggerEL97}: a simple R-tree packing method based on tiling the data space into vertical or horizontal slices recursively to construct an R-tree.
  \item  Cost-based unbalanced R-trees (\cur) \cite{DBLP:conf/ssdbm/RossSS01}: a query-aware unbalanced R-tree construction algorithm which places nodes higher in the tree based on the expected number of access under a given workload.
  \item Flood (\flood) \cite{DBLP:conf/sigmod/NathanDAK20}: a grid-based index built to efficiently process range queries by optimizing the grid structure based on the estimated query processing cost.
  \item QUery-Aware Spatial Incremental Index (\quasii) \cite{DBLP:conf/edbt/PavlovicSHA18}: a query-aware index which applies cracking \cite{DBLP:conf/cidr/IdreosKM07} during query processing to adapt the index to the query workload.
  \item Base Z-Index (\base): the base \zindex built by partitioning the data points using the median along each axis and ``\abcd'' ordering of children at each level as presented in Section~\ref{sec:basevariant}.
\end{enumerate}

\begin{figure}[t]
  \centering
  \setlength{\tabcolsep}{2.5pt}  
  \includegraphics[width=.96\linewidth]{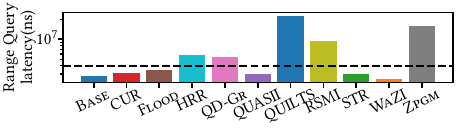}
  \caption{Average range query performance of all indexes considered in the experiments.}
  \label{fig:baselines_cutoff}
\end{figure}

We also implemented the greedy variant of Qd-tree (\qdtreegreedy) \cite{DBLP:conf/sigmod/YangCWGLMLKA20} as the RL-variant requires a large action space which is infeasible in our setting.
The cost formulation of \qdtreegreedy creates unbalanced trees tailored for disk-based indexes.
Additionally, we experimented with \zpgm\cite{DBLP:journals/pvldb/FerraginaV20,tropf1981multidimensional}, \hrr \cite{DBLP:journals/pvldb/QiTCZ18,DBLP:journals/tods/QiTCZ20}, \quilts \cite{Nishimura2017QUILTSMD} and \rsmi \cite{DBLP:journals/pvldb/QiLJK20}.
However, these five indexes performed significantly worse than the other baselines considered (as shown in Figure~\ref{fig:baselines_cutoff}).
Therefore, we chose to not include them in more detailed experimentation later.
Interestingly, four out of the five indexes we discarded (\zpgm, \hrr, \quilts, and \rsmi) are based on Z-order curves within the rank space.
Additionally, we also attempted to use \bmtree~\cite{bmtree2023} and \lmsfc~\cite{DBLP:journals/corr/abs-2304-12635} as baselines in our experiments.
However, \bmtree construction failed to finish in a reasonable duration and \lmsfc does not have publicly available implementation.

The indexes considered show a balanced representation along three different categorizations, as presented in Table~\ref{tab:baselines}.
The `SFC-based' column indicates if the construction algorithm of the indexes utilized space filling curves.
The `Query-Aware' column indicates if the index construction tailors for a specific query workload.
Finally, the `Learned' column refers to indexes that utilize any form of learning algorithms (to model cost function, density estimates, etc.) within the index or during construction.

\begin{table}[t]
  \centering
  \caption{Key properties of indexes in the experiments.}
  \label{tab:baselines}
  \begin{tabular}{cccc}
    \toprule
    Index & SFC-based & Query-Aware & Learned \\
    \midrule
    \str    &   \redcross &   \redcross & \redcross \\
    \cur     &   \redcross &   \greentick & \redcross \\
    \flood    &   \redcross &   \greentick & \greentick \\
    \quasii   &   \redcross  &   \greentick & \redcross \\
    \base    &   \greentick & \redcross & \redcross \\
    \waz  &   \greentick & \greentick & \greentick \\
    \bottomrule
  \end{tabular}
\end{table}

\spara{Implementation.}
\base and \waz are constructed with specific algorithms for finding their respective partition point and ordering.
The \base version utilizes the naive method of comparing bounding boxes to decide if a given page overlaps with the range query before filtering points within the page.
Whereas \waz utilizes the efficient skipping mechanism mentioned in Section~\ref{sec:skipping}.
We implement the two variants of \zindex in C++.\footnote{Our implementation is published at \coderepo.}

Similarly, we also implement \str \cite{DBLP:conf/icde/LeuteneggerEL97} and \quasii \cite{DBLP:conf/edbt/PavlovicSHA18} as described by the original authors.
We utilize a converged \quasii index for our evaluations, i.e., an index that has completely adapted to the query distribution without the need for further incremental updates.
A simplified \textsc{Flood}-Index for \twodim indexing (\flood) is implemented in C++, where we identify the optimal grid structure by evaluating the performance of candidate grid partitions on a sub-sample of range queries.
We adapted the \cur tree index to point data by using a `weighted' RFDE estimator to hold a weighted sum of points covering each node.
Each point is weighted by the number of distinct queries fetching the point.
Finally, we use weighted density estimates to select partitions following the Sort Tile Recursive algorithm to construct \cur.

We conducted all the experiments on a server with an Intel\textsuperscript{\textregistered} Xeon\textsuperscript{\textregistered} E5-2680v4 CPU @ 2.40GHz and 16 GB memory.
The experiments were performed on a single thread without multithreading or GPUs.
All methods were compiled with the -O3 optimizer flag.
Finally, we set the leaf node capacity \leafsize to 256 in our experiments.

\subsection{Datasets and Query Workloads}

\begin{figure}[t]
  \centering
  \setlength{\tabcolsep}{2.5pt}
  \begin{tabular}[width=\linewidth]{cccc}
  \subfloat[$\ddistr$\textsubscript{\calinev}]{\label{fig:calinev}\includegraphics[width = 0.23\linewidth,clip,trim={5mm 5mm 5mm 5mm}]{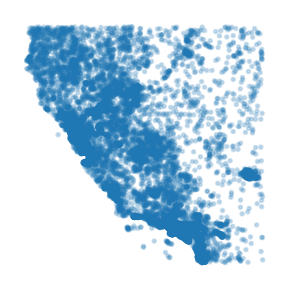}} &
  \subfloat[$\ddistr$\textsubscript{\nycity}]{\label{fig:nycity}\includegraphics[width = 0.23\linewidth,clip,trim={5mm 5mm 5mm 5mm}]{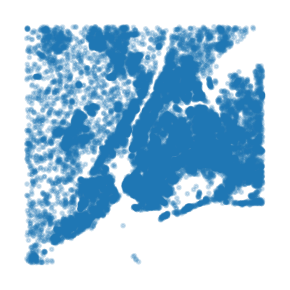}} &
  \subfloat[$\ddistr$\textsubscript{\japan}]{\label{fig:japan}\includegraphics[width = 0.23\linewidth,clip,trim={5mm 5mm 5mm 5mm}]{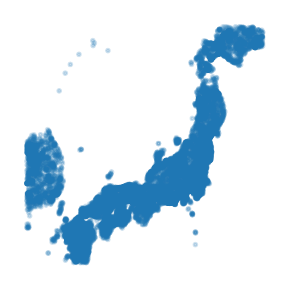}} &
  \subfloat[$\ddistr$\textsubscript{\iberia}]{\label{fig:iberia}\includegraphics[width = 0.23\linewidth,clip,trim={5mm 5mm 5mm 5mm}]{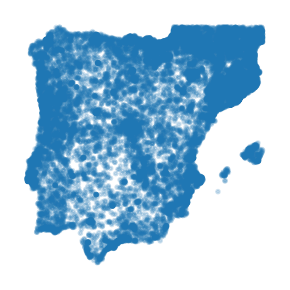}}\\
  \subfloat[$\qdistr$\textsubscript{\calinev}]{\includegraphics[width = 0.23\linewidth,clip,trim={5mm 5mm 5mm 5mm}]{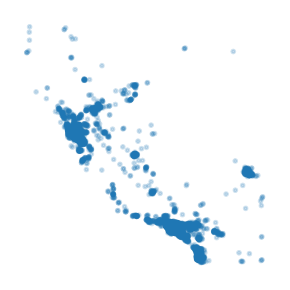}} &
  \subfloat[$\qdistr$\textsubscript{\nycity}]{\includegraphics[width = 0.23\linewidth,clip,trim={5mm 5mm 5mm 5mm}]{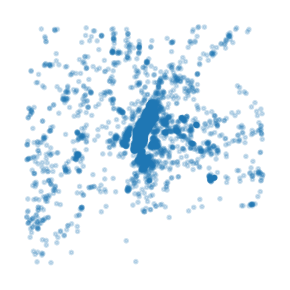}} &
  \subfloat[$\qdistr$\textsubscript{\japan}]{\includegraphics[width = 0.23\linewidth,clip,trim={5mm 5mm 5mm 5mm}]{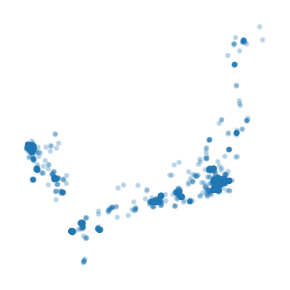}} &
  \subfloat[$\qdistr$\textsubscript{\iberia}]{\includegraphics[width = 0.23\linewidth,clip,trim={5mm 5mm 5mm 5mm}]{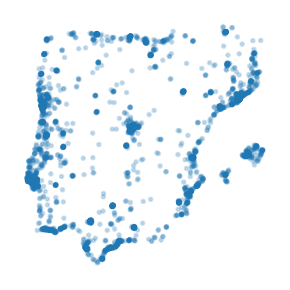}}
  \end{tabular}
  \caption{Datasets and query workloads in the experiments.}
  \label{fig:realworlddatadist}
\end{figure}

We utilize real-world datasets along with skewed semi-synthetic query workloads.
The real-world datasets consist of points of interest (POI) from OpenStreetMap~\cite{OpenStreetMap} for selected regions.
The four regions we chose are: California Coast (\calinev), New York City (\nycity), Japan (\japan), and Iberian Peninsula (\iberia).
We extract all POIs from the regions and sample these to produce datasets of appropriate size when required.
The distribution of data points \ddistr are presented in Figures~\ref{fig:calinev}--\ref{fig:iberia}.
We focus on skewed query workloads varying from the underlying data distribution for insightful analysis.
This setup differs from past works showcasing experiments that use query distributions that follow the data distribution \cite{DBLP:journals/pvldb/QiLJK20}. 
Alternatively, we present results on query workloads with real-world inspired skewness on the underlying data distribution.
We generate semi-synthetic query workloads for the real-world datasets by utilizing check-in information in the Gowalla dataset \cite{snapnets}.
Specifically, we extract check-ins that lie within the corresponding region and utilize these locations to generate a query workload resembling the distribution of check-in locations.
The process for generating range queries proceeds by sampling the centers of query rectangles from the set of check-in locations and growing along the four directions such that the query covers a portion of data space equivalent to the required selectivity.
We represent selectivity as a percentage of data space.
Figure~\ref{fig:realworlddatadist} presents the query distribution \qdistr (i.e., the check-in locations) for each of the real-world datasets.
The distributions of check-in locations are skewed towards popular locations compared to the underlying data distribution \ddistr. 
We experiment with datasets ranging from 4 to 64 million data points and range query workloads of varying selectivity (ranging from 0.0004\% to  0.1024\% of data space).
The range of parameters used for data generation are presented in Table~\ref{tab:expsetup:dataset_params}.

\begin{table}[t]
  \centering
  \setlength{\tabcolsep}{4pt}
  \caption{Parameter setting.}
  \label{tab:expsetup:dataset_params}
  \begin{tabular}{cc}
    \toprule
    Parameter &  Values (default in bold) \\
    \midrule
    Dataset size ($\times 10^6$)       &     [4, 8, 16, \textbf{32}, 64] \\
    Query selectivity (\%) & [0.0016, 0.0064, \textbf{0.0256}, 0.1024] \\
    Leaf-node size & 256 \\
    Range-query workload size & 20,000 \\
    \bottomrule
  \end{tabular}
\end{table}

\subsection{Range Query Performance}

\begin{figure*}[t]
  \centering
  \includegraphics[width=\linewidth]{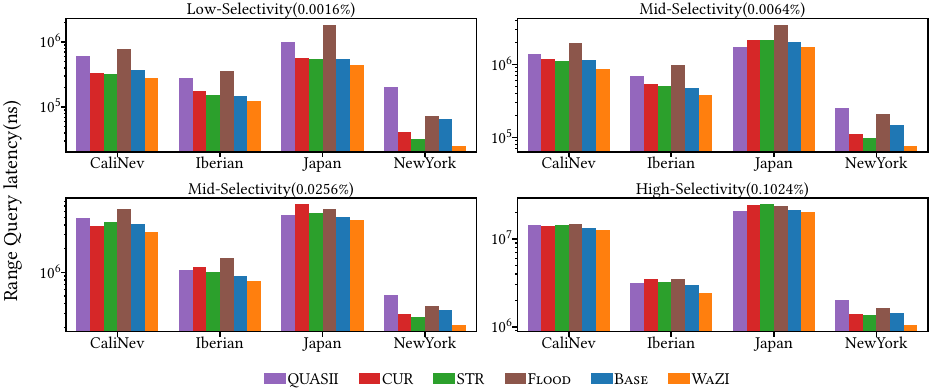}
  \caption{Average range query latency for all the indexes over four selectivity ranges. We observe that \waz consistently outperforms all the competing baselines.}
  \label{fig:real_rangequery_selectivity}
\end{figure*}

Figure~\ref{fig:real_rangequery_selectivity} presents the average range query latencies of different indexes at four levels of query selectivities (low to high).
These index latencies are evaluated over range-query workloads of 20,000 queries.
We observe that \waz clearly outperforms all other indexes in most cases.
The closest competition for \waz is from \quasii on the \japan dataset for the selectivity values 0.0064\% (top-right in Figure~\ref*{fig:real_rangequery_selectivity}) and 0.1024\% (bottom-right).
Nevertheless, \waz always performs similarly or (often substantially) better than \quasii. 
The improvements are most noticeable for the lower selectivity queries where \waz is better by about $2.3 \times$ to $8.1 \times$ over \quasii.
We also note that the indexes showcase different scales of range query latencies over different datasets.

\begin{figure}[t]
  \centering
  \includegraphics[width=0.9\linewidth]{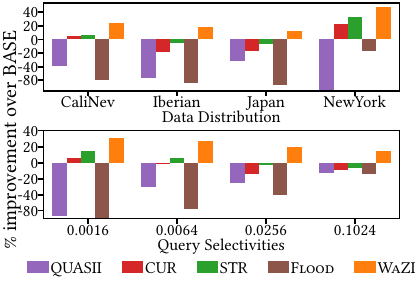}
  \caption{Percentage improvements over \base across the different data distributions and selectivities (plotted on the symlog scale with a threshold of 50.)}
  \label{fig:range_improvements}
\end{figure}
To further analyze the advantages of \waz over \base, we present a percentage improvement plot in Figure~\ref*{fig:range_improvements}.
The percentages are calculated in comparison with the performance of \base for each dataset-workload pair.
We present the average improvements of indexes over different datasets and selectivities.
Across different datasets, \waz showcases improvements between 12.8\% for the \japan dataset and 47.6\% for the \nycity dataset (top in Figure~\ref*{fig:range_improvements}).
\cur and \str showcase improvements over \base for two out of the four datasets.
We also see that the percentage differences between indexes decrease as the selectivity increases. 
The improvements by \waz linearly decrease from 31\% for 0.0016\% selectivity queries to 14\% for 0.1024\% selectivity queries.
This indicates that, as the selectivity increases, query processing encounters fewer false positives relative to the actual query result size. 
Therefore, a learned and workload-aware index is more suited for settings with low-selectivity queries.
Finally, \waz is the only index that shows a consistent improvement over \base.

\begin{figure}[t]
\centering
\includegraphics[width=0.8\linewidth]{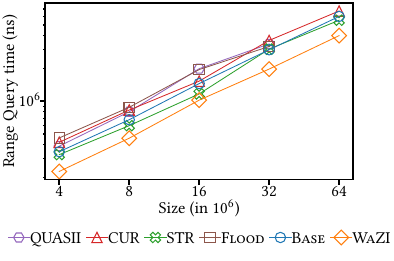}
\caption{Range query time by varying the dataset size from 4 to 64 million for a mid-selectivity (0.0256\%) query workload.}
\label{fig:range_datasize}
\end{figure}

\spara{Range Query over Different Dataset Sizes.}
Additionally, we plot the average range query latencies in Figure~\ref{fig:range_datasize} to analyze the performance of different indexes across varying dataset sizes.
The query latencies are recorded for a query workload with a selectivity of 0.0256\%.
We observe that the performance of all indexes scales nearly linearly with the dataset size, with \waz outperforming all competing baselines over different data sizes.
Furthermore, the difference in performance between the competitors and \waz remains somewhat constant with an increasing dataset size.

\begin{figure}[t]
  \centering
  \includegraphics[width=0.8\linewidth]{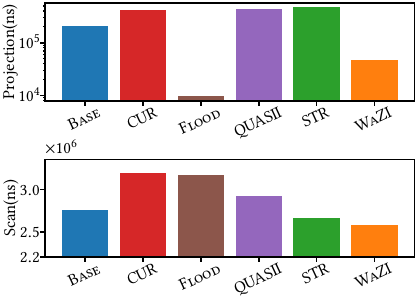}
  \caption{Range query latencies of different indexes split into the Projection and Scan phases.}
  \label{fig:projection_scan}
\end{figure}

\spara{Projection vs Scanning.}
It is useful to analyze the query processing latency of an index in two parts.
In the first part, termed Projection, the index traverses its search structure to identify all the leaf nodes (and hence the pages) overlapping a given range query. 
Second, we scan the data points within the projected pages to filter the result set.
Such analysis separates the performance overhead of the internal search structure from the performance overheads originating from the data layout of the constructed index.
Figure~\ref{fig:projection_scan} shows the two parts of query time for a 32 million-sized dataset under a mid-selectivity query workload.
In terms of projection (top in Figure~\ref{fig:projection_scan}), \cur, \quasii, and \str perform worst due to its tree-traversal cost, followed by Z-indexes \base and \waz.
\waz performs $4.3\times$ faster projection than \base.
This speedup could mostly be attributed to the efficient skipping mechanism.
\flood performs the fastest projection (about $20\times$ faster than \base) of the query to pages, as it does not perform a tree traversal for projection like the other indexes.
The scan phase of the indexing dominates the performance of an index. 
\waz outperforms other indexes in terms of scan time due to its optimized data layout.

\spara{Remark on kNN and Spatial-Join Queries.}
For spatial indexes that are not specialized to kNN or spatial-join queries (as is the case for all indexes we evaluate), kNN and spatial-join queries are typically decomposed to and processed as sets of range queries~\cite{knnspatialrange}.
As a result, query performance exhibits the same behavior as their range query performance presented above, as we have confirmed experimentally.
For this reason, we do not include additional results for kNN and spatial-join queries.

\subsection{Point Query Performance}

\begin{figure}[t]
  \centering
  \setlength{\tabcolsep}{2.5pt}  
  \includegraphics[width=0.8\linewidth]{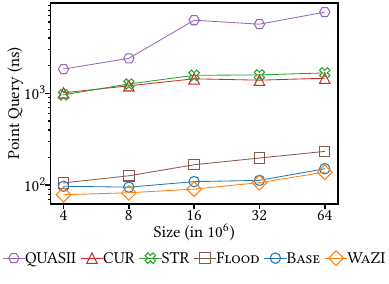}
  \caption{Point query time by varying the dataset size from 4 to 64 million. We observe that \waz performs the best across all dataset sizes.}
  \label{fig:point_datasize}
\end{figure}

Although we focus on optimizing indexes for range-query workloads, we also show the performance of proposed indexes for point queries for completeness.
We sample 50,000 point queries from the data distribution \ddistr of the respective datasets.
Figure~\ref{fig:point_datasize} presents the average query time (over different datasets) for various indexes against increasing data size.
Comparing indexes from best to worst, we see that \waz and \base outperform all other baselines.
This performance is due to simpler computations required at the internal nodes of a \zindex (see Algorithm~\ref{algo:treetraversal}) to identify the relevant child nodes.
\flood index performs similarly but slower than \waz and \base.
\str and \cur have non-overlapping child nodes and perform marginally faster than \cur which allows for overlapping.
Finally, \quasii suffers computationally due to the heavily fractured data layout caused by database cracking-based index construction.
Generally, the performance of each index over point queries is highly correlated with its respective cost of tree traversal and the underlying data layout.
We see that \waz performs better than the baselines by a factor of $1.5$--$15 \times$ due to the efficiency of node-level computations.

\subsection{Build Time}

We record the build time for all the indexes we compare in Table~\ref{tab:realexp:build_time}.
For all indexes, the build time grows linearly as the dataset size increases.
\str takes the least time for index construction as the Sort-tile recursive algorithm has a log-linear complexity.
Among the query-aware indexes (see Table~\ref{tab:baselines}), \waz performs third best behind \flood and \cur.
On comparison against learned indexes, \waz performs second best behind \flood.
\waz requires $5.4\times$ more build time than \flood.
We mainly attribute this to the need for constructing and evaluating data density estimators required for cost computations.
The build-time of \waz is similar to the query-aware \cur index which utilizes query selectivity estimation of data points to compute an approximate cost.
Therefore, compared to the baselines considered, \waz has a reasonable build time.

\begin{table}[t]
  \centering
  \setlength{\tabcolsep}{4pt}
  \caption{Build time (in seconds) of all indexes compared.}
  \label{tab:realexp:build_time}
  \begin{tabular}{crrrrrrrrrr}
    \toprule
    Size ($\times 10^6$) &   \base &    \cur & \flood  &   \quasii  &   \str &   \waz  \\
    \midrule
    4          &    7.4 &   23.8 &   4.6  &    123.6  &   1.8 &   25.1  \\
    8          &   16.2 &   48.5 &  10.1  &    418.1  &   4.0 &   51.8  \\
    16         &   35.0 &   97.0 &  25.8  &   4400.7  &   8.4 &  105.3  \\
    32         &   71.6 &  188.5 &  42.1  &  15781.5  &  19.1 &  203.2  \\
    64         &  154.6 &  366.9 &  97.3 &   57943.8  &  40.5 &  414.3  \\
    \bottomrule
    \end{tabular}
\end{table}

\spara{Cost-redemption}. 
Typically, learned indexes achieve improved query execution time, at the cost of computationally expensive index building.
To quantify this trade-off for a given dataset and query workload, we define \emph{cost redemption} as the number of query executions for which the cumulative (building+querying) time of an index is equal to the respective cumulative time of \base.
Specifically, the cost-redemption value of index \cellX is defined as $\costredemp{\cellX} = \frac{\cellX.Build-\base.Build}{\base.Query-\cellX.Query}$.
%


\begin{table}[t]
  \centering
  \setlength{\tabcolsep}{2.7pt}
  \caption{Cost-redemption value of indexes against \base. Lower is better. Missing values indicate that the index does not redeem its high-construction time with efficient query performance compared to \base. }
  \label{tab:realexp:index_redemption}
  \begin{tabular}{cccccc}
    \toprule
    Data Dist. &   \cur &    \flood &   \quasii &  \str &   \waz \\
    \midrule
\calinev    & $(+)\ 550$k  &  $(-)\ \ 13$k & $(-)$ &  $(-)\ 320$k &  $(+)\ 192$k \\
\iberia     & $(-)$                    &  $(-)\ \ 44$k  & $(-)$  &  $(-)\ 450$k &  $(+)\ 746$k \\
\japan      & $(-)$                    &   $(-)\ \ 6.5$k & $(-)$  &  $(-)\ \ 69$k &  $(+)\ 269$k \\
\nycity     & $(+)\ 1300$k &  $(-)\ 830$k  & $(-)$ & $(+)$                  &  $(+)\ 743$k \\
    \bottomrule
  \end{tabular}
\end{table}

We showcase the {cost-redemption} of the considered indexes in Table~\ref{tab:realexp:index_redemption}.
Cells marked with $(+)$ correspond to cases where the index eventually redeems its higher building time after the reported number of queries.
By contrast, cells marked with $(-)$ correspond to the opposite behavior: they have faster building time than \base, but larger query execution time, leading to larger cumulative time after the reported number of queries. 
Finally, a missing value indicates that the index has better $(+)$ or worse $(-)$ cumulative time regardless of the number of queries executed.
From Table~\ref{tab:realexp:index_redemption} we can see that \flood and \str are better than \base in the short run due to their smaller build times.
\cur is better than \base in the longer run for \calinev and \nycity datasets while being clearly worse for \iberia and \japan.
The massive build time of \quasii and the slower query performance indicate that it is the worst-performing index.
Finally, \waz is a better choice than \base if the expected query workload has at least $10^6$ queries.
Therefore, this is a limitation of \waz as it is not suited for workflows that require quick index construction to execute a few queries as found in data analysis workflows.
Instead, it is suited for workflows where index construction can be performed offline using more computational resources and deployed for an extended amount of time.

\subsection{Index Size}

%
We present the size of each index built for the datasets with 32 million points in Table~\ref{tab:realexp:index_size}.
%
%
\quilts, \flood and \quasii exhibit $0.31-0.59\times$ smaller index sizes than \waz.
The index size of \waz is similar to the R-tree-based indexes, \cur and \str.
Crucially note that the size of \waz is nearly identical to \base.
This indicates that the methods presented in this paper construct a workload-aware version of \zindex at no additional space cost.
Furthermore, we see that the size of all the indexes we utilized in our experiment grows linearly with the dataset size.

\begin{table}[t]
  \centering
  \setlength{\tabcolsep}{3.5pt}
  \caption{Sizes (in MBs) of all indexes compared.}
  \label{tab:realexp:index_size}
  \begin{tabular}{crrrrrr}
    \toprule
    Size ($\times10^6$) &   \base &    \cur & \flood &   \quasii &   \str &   \waz \\
    \midrule
    4          &    8.27 &   10.33 &   5.39 &    3.02 &   7.09 &    8.99 \\
    8          &   16.83 &   19.05 &   6.14 &    5.31 &   15.41 &   17.80 \\
    16         &   34.94 &   36.91 &  11.00 &   13.79 &   23.48 &   35.15 \\
    32         &   67.60 &   74.73 &  37.18 &   26.83 &  66.02 &   69.84 \\
    64         &  135.86 &  149.01 &  53.13 &    61.48 &  112.00 &  138.83 \\
    \bottomrule
  \end{tabular}
\end{table}

\subsection{Index Update}

\waz supports index updates like other tree-based indexes.
An index update operation in \waz is performed similarly to point query processing.
Inserting/deleting a point \querypoint proceeds by finding the leaf node enclosing \querypoint (using Algorithm~\ref{algo:treetraversal}) and updating the corresponding page.
When any page overflows/underflows, we perform a node splitting/merging operation as for any tree index.
%

\begin{figure}[t]
  \centering
  \includegraphics[width=0.75\linewidth]{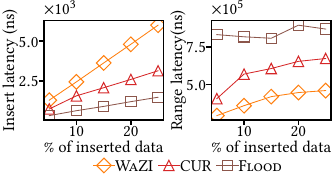}
  \caption{Insert and range query latencies over inserts.}
  \label{fig:insert_queries}
\end{figure}

%
Figure~\ref{fig:insert_queries} examines the impact of insert queries on a subset of indexes each containing 32 million data points.
We sample 8 million insert query points uniformly from the data space and incrementally insert these points in five equal-sized iterations.
For each iteration, we recorded the latencies of insert queries and range queries after insertions.
We split any overflowing pages of \waz along the data medians.
We see that \waz performs slowest in terms of insert latencies in comparison with \cur and \flood (left in Figure~\ref{fig:insert_queries}).
This can be attributed to the costly update to \pagelist during node splits and the need to recompute the look-ahead pointers.
We also note that the inserts have a minor effect on range query performance, with latencies degrading logarithmically with inserts.

\subsection{Effect of Workload Changes}
Workload-aware indexes are susceptible to degrading query performance due to the changes in workloads.
In this subsection, we compare the effects of query workload changes on \base and \waz. 
Specifically, we build \base and \waz using their original query workload and evaluate them over iteratively changing workloads.
We experiment with both uniform and skewed workload changes. 
In the former case, we replace the dataset's original workload with uniformly sampled queries over the data space.
In the latter case, we replace the dataset's original workload with the (skewed) workloads of other datasets.

\begin{figure}[t]
  \includegraphics[width=0.95\linewidth]{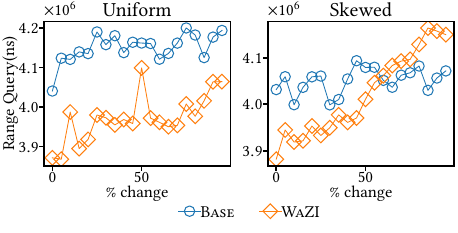}
  \caption{Range query time of \waz and \base over the changes in workloads. On the left, higher \% change corresponds to a more uniform workload; on the right, to a workload more similar to a different skewed workload.}
  \label{fig:workloadshift}
\end{figure}

The results of the experiment are presented in Figure~\ref{fig:workloadshift}.
We see that the performance of \base remains relatively stable across the range of workload changes, while \waz degrades in performance.
As the workload changes to a uniform workload (left in Figure~\ref{fig:workloadshift}), \waz degrades gracefully and remains better than \base.
This shows that \waz's skipping mechanism and the layout which is tailored to both the data and the original query workload (see Eq.~\ref{eq:greedy_totalcost}) retains some advantage over the \base index even over uniform queries.
The weakness of \waz is highlighted in the case where the workload changes to a differently skewed workload (right in Figure~\ref{fig:workloadshift}).
The performance of \waz degrades rapidly and becomes slower than \base after the workload changes more than 60\%.
This indicates that \waz is most suited for the settings with known query workloads which do not change drastically and it requires to be rebuilt if the workload changes sufficiently.

To our knowledge, automatically deciding when to rebuild a spatial index in case of workload drift is an open problem.
However, there have been extensive studies on detecting concept drift for ML models \cite{gama2014survey}.
We suggest that similar methods could be adapted for rebuilding spatial indexes.
However, the exploration of such methods is beyond the scope of the paper. 

\subsection{Ablation Study}

In this paper, we presented methods to perform adaptive partitioning and ordering (Section~\ref{sec:wazi_index:adapt}) and a novel skipping algorithm (Section~\ref{sec:skipping}) to create the \waz index.
Here we perform an ablation study for the two methods we presented. 
To facilitate comparison of each individual method mentioned above, we compare the \base and \waz index against two additional methods:
\begin{itemize}
  \item \baseplus: A \base index variant with default partitioning upon which we construct and utilize the look-ahead pointer logic.
  \item \wazminus: A \waz variant with adaptive partitioning, but no look-ahead pointers.
\end{itemize}

\begin{figure}[t]
  \centering
  \includegraphics[width =\linewidth]{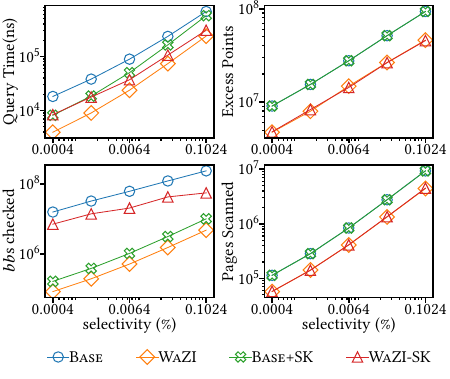}
  \caption{Results for ablation study regarding query time (top-left), excess retrievals (top-right), bounding boxes checked (bottom-left), and pages scanned (bottom-right).}
  \label{fig:zindex_comparison}
\end{figure}

We perform experiments on a dataset of size 8 million and log metrics of interest.
The results of the experiment are presented in Figure~\ref{fig:zindex_comparison}.
%
%
First, we show the incremental improvements made against \base by the additional methods in Figure~\ref{fig:zindex_comparison} (top-left). 
We see that both \wazminus and \baseplus perform similarly for low selectivity indicating both the methods contribute equally towards the improved performance.
However, as selectivity increases, \wazminus and \baseplus diverge.
At higher selectivity workloads, \baseplus tends towards the performance of \base index while \wazminus tends towards \waz.  
This trend shows that adaptive partitioning improves performance by a larger factor than the use of look-ahead pointers at high selectivity ranges.

Second, the bottom-left plot of Figure~\ref{fig:zindex_comparison} shows the number of bounding boxes (\mbr) checked for overlap with the range query while processing leaf nodes in the $[\low :\high]$ interval.
We clearly see that the use of look-ahead pointers drastically reduces the number of bounding boxes compared.
Indexes with look-ahead pointers perform $50$--$100\times$ fewer bounding box comparisons.
Third, we show the number of excess points compared (top-right) and the number of pages scanned (bottom-right).
The underlying data-layout of an index greatly impacts the performance on these two metrics. 
We see that \base and \baseplus perform worse than \waz and \wazminus indicating that the adaptive partitioning is the key differentiating factor.
Therefore, we infer that adaptive partitioning creates data layouts that are more efficient for range query processing in a \zindex.
From these results, we can conclude that the two methods we present in this paper address two key aspects of range query performance in a \zindex.
The adaptive partitioning generates better data layouts and the look-ahead pointers facilitate quicker traversal of the layout.

\section{Conclusion}
\label{sec:conclusion}

In this paper, we presented \waz, a learned and workload-aware variant of a \zindex that jointly optimizes the storage layout and search structures for spatial indexing.
Our method adapts to data distribution and query workload to build indexes to provide better range query performance.
We began by devising a metric for optimization based on the number of data points fetched during query processing, called retrieval cost.
Then we proposed an algorithm for building a \zindex variant, \waz, which minimizes the retrieval cost.
Additionally, we designed efficient mechanisms for \waz to skip irrelevant pages accessed during range query processing.
Through extensive experiments, we demonstrated that \waz improved upon existing spatial indexes, such as \flood and \cur, in terms of range query performance for real-world datasets.
We presented a comprehensive ablation study, where we showed the relative contribution of each component towards improvements shown by \waz.
Additionally, we see that \waz performs well on point query performance while providing favorable construction time and index size tradeoffs.%

For future work, we intend to explore three aspects that are not yet covered by this paper.
First, we intend to expand the cost formulation to account for other factors of index performance beyond retrieval cost so as to fit mixed types of query workloads. 
Second, we intend to explore ways to solve for the recursive cost formulation briefly mentioned in Section~\ref{sec:wazi_index:indexconst}.
Specifically, we aim to use machine learning models to efficiently learn the retrieval cost, which currently requires a dynamic programming algorithm with a state space of $O(\numdata^4)$.
Third, we intend to develop mechanisms to decide when to retrain an index when the data distribution and query workload change.

\section*{Acknowledgments}
We would like to thank Arpit Merchant and Ananth Mahadevan for their useful suggestions regarding the presentation of experimental results.
Sachith Pai and Michael Mathioudakis are supported by University of Helsinki and Academy of Finland Projects MLDB (322046) and HPC-HD (347747).
Yanhao Wang was supported by the National Natural Science Foundation of China under grant no.~62202169.

\bibliographystyle{ACM-Reference-Format}
\bibliography{refs}

\end{document}